\documentclass[11pt, a4paper]{article}
\pdfoutput=1
\usepackage{jheppub}
 \usepackage[font=small, format=hang, margin={1cm, 1cm}]{caption}
\usepackage{hyperref}
\usepackage{amsfonts} 
\usepackage{amsmath} 
\usepackage{amssymb}
\usepackage{mathtools}
\usepackage{relsize}
\usepackage{setspace}   
\usepackage{color}  
\usepackage{dutchcal}
\usepackage{wasysym}
\usepackage{booktabs}
\usepackage[utf8,applemac]{inputenc} 
\usepackage{tensor}
\usepackage{tikz}
\usepackage{graphicx}
\usepackage{caption}
\usepackage{subcaption}
\usepackage[normalem]{ulem}
\usepackage{epstopdf}
\usepackage{comment}
\numberwithin{equation}{section}
\usepackage{dcolumn}
\usepackage{bm}

\newcommand{\bea}{\begin{eqnarray}}
\newcommand{\eea}{\end{eqnarray}}
\newcommand{\be}{\begin{equation}}
\newcommand{\ee}{\end{equation}}
\def\nn{\nonumber}

\newcommand{\cA}{\mathbcal{A}}
\newcommand{\cB}{\mathbcal{B}}
\newcommand{\cC}{\mathbcal{C}}
\newcommand{\cD}{\mathbcal{D}}
\newcommand{\cE}{\mathbcal{E}}
\newcommand{\cF}{\mathbcal{F}}
\newcommand{\cG}{\mathbcal{G}}

\newcommand{\cH}{\mathbcal{H}}
\newcommand{\cK}{\mathbcal{K}}

\newcommand{\cN}{\mathbcal{N}}

\newcommand{\beqs}{\begin{eqnarray}}
\newcommand{\eeqs}{\end{eqnarray}}

\newcommand{\mn}{{\mu\nu}}
\newcommand{\p}{\partial}

\newcommand{\de}{\delta}
\newcommand{\bx}{{\mathbf x}}

\newcommand{\br}{\boldsymbol{r}}
\newcommand{\eps}{\epsilon}
\newcommand{\sub}{\mathcal{O}(1/r)}

\hypersetup{
    colorlinks,%
    citecolor=blue,%
    filecolor=blue,%
    linkcolor=blue,%
   urlcolor=blue,
   linktoc=page
}

\title{\boldmath \Large Gravitational multipole moments from Noether charges}

\author[a]{Geoffrey Comp\`ere,}
\author[a]{Roberto Oliveri}
\author[b]{and Ali Seraj}

\affiliation[a]{Universit\'{e} Libre de Bruxelles and International Solvay Institutes,\\
CP 231, B-1050 Brussels, Belgium}
\affiliation[b]{Institute for Research in Fundamental Sciences (IPM),\\P.O.Box 19395-5531, Tehran, Iran}

\emailAdd{gcompere@ulb.ac.be}
\emailAdd{roliveri@ulb.ac.be}
\emailAdd{ali\_seraj@ipm.ir}

\abstract{We define the mass and current multipole moments for an arbitrary theory of gravity in terms of canonical Noether charges associated with specific residual transformations in canonical harmonic gauge, which we call multipole symmetries. We show that our definition exactly matches Thorne's mass and current multipole moments in Einstein gravity, which are defined in terms of metric components. For radiative configurations, the total multipole charges -- including the contributions from the source and the radiation -- are given by surface charges at spatial infinity, while the source multipole moments are naturally identified by surface integrals in the near-zone or, alternatively, from a regularization of the Noether charges at null infinity. The conservation of total multipole charges is used to derive the variation of source multipole moments in the near-zone in terms of the flux of multipole charges at null infinity.} 

\arxivnumber{1711.08806}

\begin{document} 
\maketitle
\flushbottom

\newpage
\section{Introduction}

The gravitational field surrounding a stationary body in General Relativity is entirely determined by its mass and current multipole moments, up to a diffeomorphism \cite{1980CMaPh..78...75B,1981RSPSA.376..333B,1981JMP....22.1236K,1981JMP....22.2006K}. In the wave-zone region of a non-stationary system, assuming no incoming radiation, a general asymptotically flat Einstein metric is also entirely determined by its  mass and current  source multipole moments, up to a diffeomorphism  \cite{1986RSPTA.320..379B}. The mass multipole moments are already defined in Newtonian theory (see, \emph{e.g.}, \cite{Poisson:2014}), while the current multipole moments are only defined in General Relativity \cite{Geroch:1970cd,Hansen:1974zz,Thorne:1980ru,1989JMP....30.2252F}. 
 
Multipole moments are not defined intrinsically in terms of the geometry, but require a background structure involving Minkowski spacetime and either a choice of gauge or a choice of conformal completion. For outgoing-wave linearized configurations and for stationary non-linear configurations in General Relativity, Thorne \cite{Thorne:1980ru} provided a unique and well-defined definition of source multipole moments for asymptotically flat spacetimes. His definition is stated in the so-called Asymptotically Cartesian Mass Centered (ACMC$_\infty$) gauge, also known as \emph{canonical harmonic gauge} \cite{Blanchet:1998in}, obtained from the usual \emph{de Donder gauge} after further gauge fixing. Thorne's source moments agree with the Geroch-Hansen's definition for stationary spacetimes \cite{Geroch:1970cd,Hansen:1974zz}, up to a choice of normalization, after calibrating appropriately the ambiguity in the definition of the conformal factor in the Geroch-Hansen definition \cite{1983GReGr..15..737G}. 
Independently, \emph{radiative multipole moments} are defined in Bondi gauge at null infinity from the spherical harmonic decomposition of the Bondi news tensor \cite{1962RSPSA.269...21B,1962RSPSA.270..103S}. The expression of radiative multipole moments in terms of source multipole moments can be established perturbatively to all orders in the gravitational coupling under certain hypotheses including no incoming radiation \cite{Blanchet:1986dk}. Such expressions involve time integrals, known as \emph{tails}, and further non-linear terms including the non-linear memory; see \emph{e.g.}, the review \cite{Blanchet:2013haa}. 



For a Kerr black hole, the mass multipole moments $I^{lm}$ and the current multipole moments $S^{lm}$ are completely determined by its mass $M$ and angular momentum $J=M a$. Since the Kerr black hole is axisymmetric, $I^{lm}=I_l\, \de_{m,0}$ and  $S^{lm}=S_l\, \de_{m,0}$. Since it is reflection-symmetric (\emph{i.e.}, symmetric under reflection with respect to the equatorial plane), odd mass multipole moments and even current multipole moments vanish, $I_{2l+1}= S_{2l} = 0$. The non-vanishing multipoles are given by the Geroch-Hansen formulae \cite{Geroch:1970cd,Hansen:1974zz}
\bea
I_{2l} = (-1)^{l} M a^{2l}, \qquad S_{2l+1} = (-1)^l M a^{2l+1}, \label{Kerrm}
\eea
where $I_0 = M$ is the mass, $S_1 = J$ is the angular momentum and $I_2=-J^2/M$ is the mass quadrupole moment. Since the Kerr black hole is generally accepted to be the final stationary state in General Relativity, measuring more than two multipole moments in the metric surrounding a stationary black hole is a direct test of General Relativity, or more precisely, of its no-hair theorems \cite{Collins:2004ex,Cardoso:2016ryw}. Multipole moments can be tracked from the gravitational wave spectrum emission of probes that model extreme mass ratio inspirals \cite{Ryan:1995wh}. The gravitational wave detector LISA might be able to measure the mass quadrupole moment of merging black holes with good accuracy \cite{1997PhRvD..56.1845R,2007PhRvD..75d2003B}. This would allow to constrain alternative gravitational theories that predict a different mass quadrupole moment (see \cite{Berti:2015itd} for a status report and \cite{Yagi:2012ya,Ayzenberg:2014aka,Collins:2004ex,Herdeiro:2015gia} for specific models). 

Any stationary and axisymmetric metric in General Relativity can be reconstructed from given multipole moments under certain assumptions \cite{Backdahl:2005uz,Backdahl:2006ed}. However, the multipole expansion is poorly convergent in strong-field regions, such as in the vicinity of black holes \cite{Ryan:1995wh,Collins:2004ex}. For neutron stars or other types of stars, the two sets of multipole moments are determined by the matter distribution. The multipole moments of a stationary neutron star also characterize the equation of state of the matter it consists of; see, \emph{e.g.},~\cite{1988FCPh...12..151D,1994A&A...291..155S,1999ApJ...512..282L}. Gravitational wave emission from binary neutron star mergers, such as the one recently observed \cite{GBM:2017lvd}, or black hole-neutron star mergers, therefore, contain signatures of the equation of state of neutron stars. The effect of the mass quadrupole moment is small but comparable in magnitude to spin-spin interactions \cite{1993PhRvD..47.4183K}.   

Multipole moments are generated by sources. In linearized General Relativity, multipole moments can be expressed as volume integrals depending on the source stress-tensor \cite{1971Phy....53..264C,Campbell:1977jf,Damour:1990gj}. For slow moving sources ($ v \ll c$), the post-Newtonian (PN) expansion of General Relativity is applicable and multipole moments can be expressed as volume integrals over the sources \cite{1971JMP....12..401B,1975ApJ...197..717E,Thorne:1980ru,Blanchet:1998in}. 
These volume integrals can be expressed as surface integrals in the outer near-zone, \emph{i.e.}, far from the source but at radii negligible with respect to the radiation wavelength \cite{Blanchet:2004re}. 

In this paper, we provide a  definition of \emph{mass} and \emph{current multipole moments} for an arbitrary generally covariant theory of gravity. It agrees, in the case of Einstein gravity, with Thorne's definition of source multipole moments both for outgoing-wave linearized configurations and for stationary non-linear configurations. While we allow for matter, we do not consider the additional multipole moments that arise from the presence of matter. We instead only consider the mass and current multipole moments, including their matter contribution. Our definition extends to null infinity where it extracts the source multipole moments as a function of retarded time. Radiative multipole moments, which are not derived in this paper, are functionals of these source multipole moments involving non-linearities, tails and memory.

Our programme is achieved by defining multipole moments from the canonical Noether charges associated with specific residual symmetries of canonical harmonic gauge. First, we will characterize the infinitesimal form of three sets of \emph{large coordinate transformations} that preserve the canonical harmonic gauge and that generalize the Poincar\'e algebra of Killing symmetries. The vector fields generating these transformations will be denoted \emph{multipole symmetries}. Given an arbitrary generally covariant Lagrangian, one can define surface charges associated with an arbitrary vector field using either covariant phase space methods \cite{Iyer:1994ys} or cohomological methods \cite{Barnich:2001jy,Barnich:2007bf}. The explicit expressions are summarized in Appendix \ref{app:ch} for General Relativity and in \cite{Iyer:1994ys,Compere:2009dp,Azeyanagi:2009wf} for a large class of higher curvature gravity theories coupled to bosonic matter. We will provide expressions for the gravitational multipole moments in terms of the canonical multipole charges, and find precise agreement with Thorne's definitions, up to a choice of normalization. This will allow us to calibrate our definition, which then straightforwardly extends to a general theory.

The multipole symmetries are generically \emph{outer symmetries}: they are not tangent to the phase space defined by the boundary conditions (\emph{i.e.}, the Lie derivative of the metric with respect to these symmetries do not fall off sufficiently fast), except for the lowest harmonic modes which form the Poincar\'e algebra. Yet, these symmetries are associated with well-defined conserved Noether charges at spatial infinity, as we will show. 

In the radiative case, the conservation equation will be used to relate the variation of source multipole moments, evaluated in the near-zone region, to the flux of multipole charges at null infinity, by means of the \emph{generalized Noether theorem} for gauge theories \cite{Wald:1993nt,Barnich:2000zw,Barnich:2001jy}. The flux of multipole charges is not the observable radiative multipole moments encoded in the Bondi news tensor. Instead, it is the time variation of the source multipole moments which is encoded in subleading components of the metric in Bondi gauge. 

Our construction extends to gravity theories the analysis of electromagnetic multipole charges in Maxwell theory \cite{Seraj:2016jxi}. 

The rest of the paper is organized as follows. In Section \ref{sec2-Residual symmetries}, we define the canonical harmonic gauge independently of field equations and, after an overview of several concepts of symmetries, we find the multipole symmetries. In Section \ref{sec:stat}, we compute the canonical charges associated to multipole symmetries for non-linear stationary configurations in Einstein gravity and relate them to the multipole moments of the source. In Section \ref{sec:lin}, we discuss radiating geometries at the linearized level. We express the multipole moments as Noether charges and derive their conservation equation. We conclude with several avenues for further research in Section \ref{sec:discussion}. The appendices gather our conventions and computational details.

\section{Harmonic gauge and residual transformations}\label{sec2-Residual symmetries}
\subsection{Prelude on symmetries in gravity and gauge theories}
It is by now a well know fact that gauge transformations exhibit novel features on manifolds with (asymptotic or finite) boundaries. This effect is sometimes interpreted as the spontaneous breaking of gauge invariance by the boundary conditions. More precisely, while the local gauge transformations are degeneracies of the presymplectic form of the theory and are hence quotiented out in the physical phase space \cite{Lee:1990nz}, those with support at the boundary are not degeneracies and form vector fields acting on the phase space. These ``asymptotic symmetries'' are required to generate Hamiltonian vector fields on the phase space whose generators are finite and conserved quantities. In other words, the boundary conditions require that all canonical charges associated with asymptotic symmetries be finite, conserved and integrable in the sense of \cite{Regge:1974zd,Wald:1999wa,Barnich:2007bf}. When these conditions are met, one obtains, after quotienting by local gauge symmetries, the group of asymptotic symmetries, \emph{i.e.}  
\begin{align*}
\text{Asymptotic symmetry group} = \frac{\text{Boundary conditions preserving gauge transformations}}{\text{Local gauge transformations}}.
\end{align*}
For asymptotically flat spacetimes, the asymptotic symmetry group at spatial infinity is the BMS group, consisting of supertranslations and Lorentz transformations, which contains the Poincar\'e group as a subgroup \cite{Arnowitt:1959ah,Regge:1974zd,1962RSPSA.269...21B,1962RSPSA.270..103S,Ashtekar:1978zz,Blanchet:1992br,Christodoulou:1993uv,Barnich:2009se,Compere:2011ve,Barnich:2011mi,Virmani:2011aa,Strominger:2013jfa,Troessaert:2017jcm}:
\begin{itemize}
    \item Asymptotic symmetries $\simeq$ BMS symmetries $\supset$ Poincar\'e symmetries.
\end{itemize}
Now, one may find Hamiltonian vector fields that are \textit{not} tangent to the phase space, but still associated with finite, conserved and integrable canonical charges on the phase space. We call \emph{phase space symmetries} the coordinate transformations that are associated with finite, conserved and integrable canonical charges on the phase space. The phase space symmetries contain the asymptotic symmetries and other symmetries that violate the boundary conditions, which we call \emph{outer symmetries}:
\begin{align*} 
\text{Phase space symmetries} = \text{Asymptotic symmetries } \cup \text{ Outer symmetries}.
\end{align*}
In geometrical terms, the boundary conditions impose a set of constraints which determine the phase space as a submanifold in the space of field configurations (further quotiented by local gauge transformations). The asymptotic symmetries then generate vector fields tangent to this constraint submanifold and their corresponding canonical charges are functions on the phase space. However, there is still the possibility of Hamiltonian vector fields that are not tangent to the phase space, but such that their canonical charges \textit{are} functions on the phase space. This is another way to say that outer symmetries violate the boundary conditions, but are associated with finite and integrable charges. 

We will not attempt at a rigorous analysis of the mathematical framework for describing outer symmetries. Instead, we will discuss how this concept is useful in gravity: we will show that the two sets of multipole moments of Einstein gravity are precisely associated with particular phase space symmetries. We call these \emph{multipole symmetries} which form a natural extension of the Poincar\'e generators, distinct from the BMS symmetries:  
\begin{itemize}
\item Phase space symmetries $\supset$ Multipole symmetries  $\supset$ Poincar\'e symmetries.
\end{itemize}
Since the multipolar structure of a metric configuration defines a hierachical subleading structure at spatial infinity, the multipole symmetries form in compensation a fine-tuned overleading hierachy, such that the resulting canonical charges are finite. The lowest multipolar modes are the Poincar\'e symmetries, while the higher modes are outer symmetries that violate the boundary conditions.

Several other examples of outer symmetries have recently appeared in electromagnetism \cite{Campiglia:2016hvg,Seraj:2016jxi} and in gravity \cite{Compere:2016jwb,Mirbabayi:2016xvc,Conde:2016csj} (see also an alternative perspective in~\cite{Conde:2016rom,Mao:2017tey}).

\subsection{Residual transformations}
We consider an arbitrary theory of gravity in $3+1$ dimensions with dynamical metric $g_{\mu\nu}$.  We denote the inverse metric as $g^{\mu\nu}$ and the Minkowski metric as $\eta_{\mu\nu}$, whose inverse is $\eta^{\mu\nu}$. We define the field ${\mathfrak g}^{\mu\nu} \equiv \eta^{\mu\nu} -\sqrt{-g} g^{\mu\nu} $ and impose the \emph{de Donder} or \emph{harmonic gauge}
\begin{align}\label{deDonder gauge}
\partial_{\mu} {\mathfrak g}^{\mu\nu} = 0. 
\end{align}
The exact equations of General Relativity are written in this gauge as (see, \emph{e.g.},~\cite{Blanchet:2013haa})
\begin{equation}
\square_{\eta} {\mathfrak g}^{\mu\nu} = - 16 \pi \tau^{\mu\nu} \equiv 16 \pi |g| T^{\mu\nu} + \Lambda^{\mu\nu},
\end{equation}
where $\Lambda^{\mu\nu}$ is the effective stress tensor of the gravitational field and is quadratic or higher order in powers of ${\mathfrak g}^{\mu\nu}$ and $\square_{\eta}=\eta^{\mu\nu}\p_\mu \p_\nu = -\partial^{2}_t + \bar\nabla^2$ is the D'Alembertian operator with respect to the background flat metric.
Due to the de Donder gauge, the stress-energy \emph{pseudo-tensor} $\tau^{\mu\nu}$ is conserved $\partial_{\mu} \tau^{\mu\nu} = 0$.

As we have fixed the harmonic gauge \eqref{deDonder gauge}, the gauge transformations of the theory reduce to the \emph{residual transformations} of harmonic gauge. It can be shown that an infinitesimal diffeomorphism $x^\mu \to x^\mu +\xi^\mu $ respecting harmonic gauge \eqref{deDonder gauge} solves the equation 
\begin{align}\label{residual vector}
\square_g \xi^\mu=0.
\end{align}

We  further restrict the phase space and, accordingly, the harmonic vector fields~\eqref{residual vector} to those preserving the following asymptotic behavior of the lapse and shift at spatial infinity
\begin{align}\label{BC}
g_{0\mu}= \eta_{0\mu} + \mathcal{O}(1/r). 
\end{align}
However, since we are considering outer symmetries, we do \emph{not} enforce the vector fields to preserve the boundary conditions $g_{ij} =  \mathcal{O}(1/r)$ on the spatial components. At this step of our procedure, we break Lorentz covariance by explicitly choosing an asymptotic time foliation and therefore a notion of asymptotic observer at rest. Our reasoning herebelow will not depend upon the boundary conditions at future or past null infinity (see later on Eq.~\eqref{CK-condition}).

Setting $\mu=0$ in \eqref{BC} implies that $\p_0 \xi^0=\sub$, hence 
\begin{align}
\xi^{0}&=\eps(\bx)+\sub .
\end{align}
Setting $\mu=i$ in \eqref{BC}, we find 
\begin{align}
\xi^\mu=\Big(\eps(\bx)\,,\, \chi^i(\bx) +t\,\eta^{ij}\p_j \eps(\bx)\Big)+\sub.
\end{align}
The de Donder gauge condition \eqref{deDonder gauge} now amounts to  
\begin{align}
\nabla^2\eps(\bx)=0,\qquad \nabla^2 \chi_i (\bx) =0,
\end{align}
where $\nabla^2 = \eta^{ij} \p_{i} \p_{j}$ is the spatial Laplacian operator. The harmonic function $\eps$ admits two branches of solutions: either of the form $r^{-(l+1)} Y^{l m}(\theta,\phi)$ or $r^{l} Y^{l m}(\theta,\phi)$ in terms of spherical harmonics.  The first set are gauge transformations that we discard. We, instead, consider the sum of \emph{regular solid scalar harmonics} $r^{l} Y^{l m}(\theta,\phi)$
\bea
\eps(r,\theta,\phi) = \sum_{l=0}^\infty \sum_{m=-l}^l \eps_{l m} r^{l} Y_{lm}(\theta,\phi),
\eea
where $\eps_{lm}$ are arbitrary coefficients. The function can also be expanded in Cartesian harmonics
\begin{align}\label{Cartesian expansion}
\eps&= \sum_{l=0}^\infty \eps_{A_l}\mathbcal X_{A_l}=\eps_0+(\eps_x x+\eps_y y+\eps_z z)+\eps_{ij}\left(x^ix^j-\frac{x^2}{3}\delta^{ij}\right)+\cdots
\end{align}
This expression defines the monopole term ($l=0$), dipole term ($l=1$), quadrupole term ($l=2$) and so on.
Our notation and conventions are relegated to Appendix \ref{sec:conv} and properties of spherical harmonics are summarized in Appendix \ref{sec-spherical harmonics}. 

The intuitive idea for defining multipole symmetries is that a vector of components of order $r^l$ will be able to probe terms of order $r^{-l}$ in the metric, and therefore, allow to define the $l$-multipole. The harmonic equation ensures that there is a correlation between the radial behavior $r^l$ of the components of the vector field and the spherical harmonic $Y^{l m}$. Thanks to the orthogonality of spherical harmonics, it will allow to single out the corresponding multipole; see also the discussion around Eq.~\eqref{proj}.

In the vector case, a general harmonic is a linear combination of three pure gauge transformations, which we discard, and three large gauge transformations of the form (see appendix \ref{genres} for details)
\begin{align}
\boldsymbol{\chi} &=\chi^i\p_i= \br \times \nabla \eps_1(\bx) + \nabla \eps_2(\bx) + \boldsymbol V(\bx) \label{decomp},
\end{align}
where $\nabla\eps=\eta^{ij}\,\p_i\eps\,\p_j$ and $\times$ denotes the cross product defined in three-dimensional flat space. The functions $ \eps_1, \eps_2$ are combinations of regular solid scalar harmonics and $\boldsymbol V$ defined in \eqref{V-vector}, is a combination of irreducible vector harmonics which cannot be expressed in terms of one harmonic scalar. However, we discard this vector since the associated charge does not contain any nontrivial information in Einstein theory, as discussed below equation \eqref{chargeV}. However, it might be relevant for a general theory of gravity, as discussed in section~\ref{sec:discussion}.

In summary, we define the \emph{multipole symmetries} as the three sets of residual transformations: 
\begin{subequations} \label{residual symm}
\begin{align} 
K_\eps&=\eps \,\p_t+t\,\nabla \eps\,,\\
L_\eps&=-  \br\times\nabla\eps\,,\label{signchoice}\\
P_\eps&=\nabla \eps\,.
\end{align}
\end{subequations}
These symmetries can be expanded in terms of scalar and vector spherical harmonics (introduced in \ref{sec-VSH}) as
\begin{subequations}
\begin{align}
K_{\eps} &=  \sum_{l=0}^\infty \sum_{m=-l}^l \eps_{l m} \left[r^{l} Y_{lm}\,\partial_t + t~r^{l-1}\left(\sqrt{l(l+1)}~^{E}Y_{l m}^{i} + l ~^{R}Y_{l m}^{i} \right)\partial_{i}\right],\\
L_{\eps} &= -\sum_{l=1}^\infty \sum_{m=-l}^l \eps_{l m}\sqrt{l(l+1)}~r^{l}~^{B}Y_{l m}^{i}\partial_{i},\\
P_{\eps} &=  \sum_{l=1}^\infty \sum_{m=-l}^l \eps_{l m}~r^{l-1}\left(\sqrt{l(l+1)}~^{E}Y_{l m}^{i} + l ~^{R}Y_{l m}^{i} \right)\partial_{i} .
\end{align}
\end{subequations}

Upon expanding in Cartesian harmonics \eqref{Cartesian expansion} or, equivalently, in \textit{real} spherical harmonics $Y_l^m$ defined in \eqref{realY}, we find the Poincar\'e algebra as the $l=0,1$ subset of harmonics of the three vectors $K_\eps$, $L_\eps$, $P_\eps$. To show this, let us expand the harmonic function $\eps$ in terms of the real spherical harmonics and define $L_{l}{}^m=L_{\eps=r^l Y_l^m}$ and similarly $K_l{}^m,\,P_l{}^m$.
Then, it can be checked that

\paragraph{Rotations}
The $SO(3)$ rotations are generated by the $(l = 1, m )$ modes of $L_{\epsilon}$ 
\begin{subequations}
\begin{align}
L_1^{\;\;1} &=  y \partial_{z} - z \partial_{y}=L_x,\\
L_1^{-1}  &= z \partial_{x} - x \partial_{z}=L_y,\\
L_1^{\;\;0} &= x \partial_{y} - y \partial_{x}=L_z,
\end{align}
\end{subequations}
where the last equality in each line refers to the Cartesian expansion \eqref{Cartesian expansion} (\emph{i.e.}, $L_x=L_{\eps=x}$), which also coincides with the standard notation for generators of rotation. We fixed the sign convention in \eqref{signchoice} so that $L_z=+\p_\phi$ with $\eps_{r\theta\phi}=+1$, which is the opposite convention than \cite{Iyer:1994ys}.

\paragraph{Boosts}
The $(l = 1, m )$ modes of $K_{\epsilon}$ give the Poincar\'e boosts
\begin{subequations}
\begin{align}
K_1^{\;\;1}&= x \partial_{t} + t \partial_{x} = K_{x} ,\\
K_1^{-1} &= y \partial_{t}  + t \partial_{y}=K_{y} ,\\
K_1^{\;\;0}&= z \partial_{t} + t \partial_{z}=K_{z} .
\end{align}
\end{subequations}

\paragraph{Translations}
The $l = 0$ mode of $K_{\epsilon}$  gives the translation in time
\bea
K_0^{\;\;0} = \partial_t,
\eea
while the $(l = 1, m )$ modes of $P_{\epsilon}$ give spatial translations
\begin{subequations}
\begin{align}
P_1^{\;\;1} &= \partial_{x}=P_{x} ,\\
P_1^{-1} &=  \partial_{y}=P_{y},\\
P_1^{\;0}  &=  \partial_{z}=P_{z}.
\end{align}
\end{subequations}

All vectors that are distinct from the standard Poincar\'e vectors are outer symmetries. Note that, contrary to the $l=0,1$ vectors, they do not generate a closed algebra under the Lie bracket. We leave as an open problem to determine whether an algebra of multipole symmetries exists using a modified bracket, along the lines of \cite{Seraj:2017rzw}. The various vectors and their physical interpretation are summarized in Table \ref{table:1}.  

The terminology of mass and current multipole symmetries is standard and will be clear once we compute the conserved charges associated with these transformations in the next sections. We will name the multipoles associated with $P_{lm}$  as the momentum multipoles. We will interpret these multipoles in Sections \ref{sec:stat} and \ref{sec:lin}. 

\begin{table}[htb!]
\centering
\begin{tabular}{ c|ccc } 
  \toprule
 \multicolumn{1}{r}{} & $l=0$ & $l=1$&$l\geq 2$ \\ [0.5ex] \midrule
 $K_{l m}$ & Time translation & Boosts& Mass multipole symmetries\\
  $L_{l m}$ & $\varnothing$ & Rotations& Current multipole symmetries \\ 
 $P_{l m}$& $\varnothing$ &Spatial translations&Momentum multipole symmetries  \\
\bottomrule
\end{tabular}
\caption{Multipole symmetries and the Poincar\'e algebra}
\label{table:1}
\end{table}

\subsection{Canonical harmonic gauge}
\label{sec:gf}

As shown by many authors, the definition of multipole moments is ambiguous in harmonic gauge. Here, we will further fix the gauge in order to uniquely fix the multipole moments. Since our aim is to derive a definition of multipole moments for \emph{any} theory of gravity, we cannot rely on properties of Einstein solutions in order to determine the gauge fixing conditions. Instead, we will only rely on the  harmonic decomposition of tensor representations of $SO(3)$. 

For that purpose, we decompose the field ${\mathfrak g}^{\mu\nu}$ in $SO(3)$ scalar, vector and tensor, respectively, ${\mathfrak g}^{00}$, ${\mathfrak g}^{0i}$, ${\mathfrak g}^{ij}$.
We only consider nonlinear metrics which can be obtained from a perturbative construction starting from a linearized metric $g_\mn=\eta_\mn+h_{\mu\nu}$. 
At the linearized level, the field ${\mathfrak g}^{\mu\nu} = \gamma^{\mu\nu}+ O(h^2)$ is the trace-reversed metric perturbation, $\gamma^{\mu\nu} = \eta^{\mu\alpha} \eta^{\nu\beta} h_{\alpha\beta}-\frac{1}{2}\eta^{\mu\nu}\eta^{\alpha\beta}h_{\alpha\beta}$. 

A gauge transformation acts as $h_{\mu\nu} \mapsto h_{\mu\nu}  + \partial_{\mu}\xi_{\nu} + \partial_{\nu}\xi_{\mu}$ and therefore as 
\bea
\gamma_{\mu\nu} \mapsto \gamma_{\mu\nu} + \partial_{\mu}\xi_{\nu} + \partial_{\nu}\xi_{\mu} - \eta_{\mu\nu} \p_\alpha \xi^\alpha,
\eea
where all indices are lowered with the flat metric. 

As explained by Thorne \cite{Thorne:1980ru}, the symmetric trace-free (STF) harmonic tensor decomposition is the most adapted to describe the multipolar structure of the gravitational field. Indeed, remember that the very definition of multipoles originates from a STF decomposition or, equivalently, a spherical harmonic decomposition of a harmonic scalar, see Eq.~\eqref{Cartesian expansion}. Conceptually, we aim to deduce the definition of some $lm$-multipole $M^{lm}$ from a projection of the metric using a vector $\xi[\eps]$ depending on a scalar harmonic $\eps=\eps_{lm}$. Schematically, 
\bea
M^{lm} \sim \langle  \xi[\eps] , g_{\mu\nu} \rangle\label{proj}.
\eea
We will define this so far abstract projection $\langle \cdot \rangle$ later on in Eq.~\eqref{Fullcharge} and in Eqs.~\eqref{Fullcharge2}-\eqref{sub} as particular integrals over the sphere. Here, we simply note that the multipole moments will be extracted from the metric thanks to the property of orthogonality of spherical harmonics or, equivalently, of the STF tensors. Therefore, it is computationally simpler to consider the STF decomposition of the metric (instead of other decompositions such as the SVT (scalar-vector-tensor) decomposition used in cosmological perturbation theory).

In Thorne's analysis, however, the linearized Einstein's equations were used. Here, we simply generalize his considerations to a general STF decomposition without resorting to field equations. The general linearized metric can be decomposed in STF tensors as (see Appendix \ref{sec:conv} for notations)
\bea
\gamma_{00} &=& \p_{A_l}\cA_{A_l}(u,r)  \nonumber \\
\gamma_{0i} &=& \p_{A_{l-1}}\cB_{i A_{l-1}}(u,r) + \p_{p A_{l-1}}\left(\epsilon_{ipq} \cC_{q A_{l-1}}(u,r)\right) + \p_{iA_{l}}\cD_{A_l}(u,r)\label{genlin}\\
\gamma_{ij} &=& \delta_{ij} \p_{A_l}\cE_{A_l}(u,r) + \p_{A_{l-2}}\cF_{i jA_{l-2}}(u,r) + \p_{pA_{l-2}}\left(\epsilon_{pq(i} \cG_{j)q A_{l-2}}(u,r)\right) +  \nonumber\\
&+& \left[\p_{j A_{l-1}}\cH_{i A_{l-1}}(u,r) + \p_{jpA_{l-1}}\left(\epsilon_{ipq} \cN_{qA_{l-1}}(u,r)\right)\right]^{S} + \p_{ijA_{l}}\cK_{A_l}(u,r),\nonumber
\eea
where all coefficients are functions of $r$ and $u=t-r$. 
The de Donder gauge fixes the constraints
\bea
\cB_{A_l} &=& \dot\cA_{A_l} - \nabla^2 \cD_{A_l}, \\
\cE_{A_l} &=& \dot\cD_{A_l} - \frac{1}{2}\cH_{A_l} - \nabla^2 \cK_{A_l},\\
\cF_{A_l} &=& \ddot\cA_{A_l} - \nabla^2 \dot\cD_{A_l} - \frac{1}{2}\nabla^2\cH_{A_l},\\
 \cG_{A_l} &=& 2\dot\cC_{A_l} - \ddot \cN_{A_l},
\eea
where dots denote time derivatives. 
For retarded harmonic fields, which therefore obey linearized Einstein's equations, all functions $\cA_{A_l}(u,r)$, $\cB_{A_l}(u,r)$, $\dots$ take the form $\frac{1}{r}\cA_{A_l}(u)$, $\frac{1}{r}\cB_{A_l}(u)$, etc. We assume that all such functions are $\mathcal{O}(1/r)$ in general as a result of asymptotically flat boundary conditions. 

A generic retarded harmonic vector field, which preserves the asymptotically flat boundary conditions, can be used to further gauge fix the de Donder gauge to the canonical harmonic gauge; see Eq.~(8.9b) of~\cite{Thorne:1980ru} and Eq.~\eqref{gauge transfs2} for its explicit expression in spherical harmonics. One can use this vector field to remove the $1/r$ components of $\cD_{A_l}(u,r), \cH_{A_l}(u,r)$, $\cN_{A_l}(u,r), \cK_{A_l}(u,r)$, which leads to 
\begin{align}\label{canonical gauge}
\{ \cD_{A_l}(u,r), \cH_{A_l}(u,r), \cN_{A_l}(u,r), \cK_{A_l}(u,r)\} =\mathcal{O}\left(\dfrac{1}{r^2}\right).
\end{align}
In Einstein gravity, these functions are then exactly zero but they may be non-zero in alternative theories of gravity. We further apply a Lorentz boost and a spatial translation to put the system in the center of mass frame, leading to $\mathcal A_i = \mathcal O(r^{-2})$. This defines the canonical harmonic gauge in the linearized theory. 

In the non-linear theory, one computes at each perturbative order the next metric perturbation up to a linearized diffeomorphism. We fix the linearized diffeomorphism ambiguity in this expansion (of the form $\partial_{\mu}\xi_{\nu} + \partial_{\nu}\xi_{\mu} - \eta_{\mu\nu} \p_\alpha \xi^\alpha$) by canceling again the same coefficients in the radial harmonic decomposition of ${\mathfrak g}^{\mu\nu}$. Non-linearities will introduce additional radial subleading terms for each STF tensor harmonic which will not be gauged fixed. Only the leading $1/r$ terms in the radial expansion of each STF tensor harmonic  will be gauge fixed. This defines the canonical harmonic gauge for the non-linear theory. 

The resulting asymptotic expansion of the non-linear field in STF tensor harmonics has been presented explicitly for Einstein gravity \cite{Thorne:1980ru,Blanchet:1985sp}.

\section{Multipole charges for stationary solutions}
\label{sec:stat}

Stationary asymptotically flat spacetimes admit a globally defined Killing vector $\xi^\mu$ which is normalized to $-1$ at spatial infinity. Its norm, which is usually denoted as $\lambda = -\xi^\mu \xi_\mu$, encodes, in canonical harmonic gauge, the mass multipole moments of the gravitational field  \cite{Geroch:1970cd,1983GReGr..15..737G}\footnote{Geroch proved it for static spacetimes. Hansen proved in the stationary case that the mass multipole moments are determined from the norm \emph{and} twist \cite{Hansen:1974zz} but, in fact, it is clear from the proof of G\"ursel that only $\lambda$ is necessary in harmonic gauge. The remark applies for the current multipole moments. The proof of G\"ursel shows that $\omega$ is enough to find them in harmonic gauge.}. In vacuum Einstein gravity, the form 
\bea
\omega_\alpha = \epsilon_{\alpha\beta \mu\nu} \xi^\beta \nabla^\mu \xi^\nu,\label{oo}
\eea 
is closed and therefore locally exact, $\omega_\mu = \p_\mu \omega$. This allows to define the twist $\omega$ that encodes, in canonical harmonic gauge, all current multipole moments \cite{1983GReGr..15..737G}. In other words, the twist is defined as the dual scalar field associated with the Kaluza-Klein vector along the Killing coordinate $t$ (defined by $\xi^\mu \p_\mu t = 0$). Moreover,  the $4d$ vacuum stationary Einstein's equations can be rewritten in terms of $3d$ Einstein's theory coupled to a $SL(2,\mathbb R)/U(1)$ coset model parameterized by the norm and twist of $\xi^\mu$ \cite{Geroch:1970nt}. In particular, the $SO(2)$ hidden symmetries transform into themselves the two scalars $\lambda$, $\omega$ which generate the two sets of multipole moments.

If one does not assume vacuum Einstein's equations, acting with a derivative on \eqref{oo} one has more generally 
\bea
\p_{[\mu} \omega_{\nu]} = -\epsilon_{\mu\nu\alpha\beta}\xi^\alpha R^\beta_{\; \gamma} \xi^\gamma .\label{do}
\eea
In the case of a stationary minimally coupled scalar ($\xi^\mu \p_\mu \phi = 0$), the following condition on the matter stress-tensor $T^{\mu\nu}$ holds
\bea
\xi^{[\alpha} T^{\beta]}_{\;\; \gamma}\xi^\gamma = 0,
\eea
 and the twist can be defined without change. The same applies in scalar-tensor theories without matter such as Brans-Dicke theory in the formulation where the metric obeys Einstein's equation coupled to a scalar field with non-trivial kinetic term. In all these theories, only two sets of multipole moments are defined in terms of $\lambda$ and $\omega$. 
 
A generalization to electrovacuum spacetimes also exists \cite{Simon:1983kz,1990CQGra...7.1819H,Sotiriou:2004ud}. Stationary Einstein-Maxwell fields are constructed from a three dimensional metric and four scalar fields which  form a $SU(2,1)/(SU(2) \times U(1))$ coset \cite{1973JMP....14..651K}. The right-hand side of Eq.~\eqref{do} can then be written as $-\p_{[\mu} \omega^A_{\nu]}$ 
 and the twist scalar is defined through $\omega_\mu + \omega_\mu^A = \p_\mu \omega$.
 Again, there are two sets of gravitational multipole moments (together with the electric and magnetic multipole moments). 
  
Now, for stationary configurations in more general theories, the right-hand side of \eqref{do} may be non-vanishing and may not be written as an exact differential. Correspondingly, the Kaluza-Klein reduction of the theory may depend upon a $3$-dimensional vector field which cannot be dualized to a scalar. In such theories, there are three types of vector harmonics of $SO(3)$ in the decomposition of $\omega_\mu$ which together will encode the multipole structure. More precisely, the four sets of harmonics encoded in $\lambda$ and $\omega_\mu$ are related since one has the equation (see, \emph{e.g.},~\cite{Brink:2013via,Chodosh:2015nma})
\bea
\nabla_\alpha \omega^\alpha = \frac{2}{\lambda}\omega^\alpha \nabla_\alpha \lambda. 
\eea 
We expect therefore three independent sets of multipoles encoded in $\lambda$, $\omega_\alpha$.

In the following, we will present our definition of mass and current multipole moments in terms of Noether charges, which are defined with respect to the vectors fields \eqref{residual symm} and the Lagrangian. We will explicit our definition only for vacuum Einstein solutions which will allow us verify the complete equivalence to Thorne's definition. This allows us to calibrate our definition. Now, given an arbitrary generally covariant Lagrangian, we propose to define the multipole moments as the Noether charges associated with the very same vector fields \eqref{residual symm}. This provides a precise canonical definition of gravitational multipole moments for any generally covariant theory. 


\subsection{Phase space in canonical harmonic gauge}

Let $g_{\mu\nu} = \eta_{\mu\nu} + \mathcal{O}(1/r)$ be an asymptotically flat stationary metric in de Donder coordinates. For Einstein metrics, the metric coefficients $g_{\mu\nu}$ are analytic functions of the de Donder coordinates \cite{1970PCPS...68..199H}. We restrict our analysis to Einstein metrics from now on. In canonical harmonic coordinates described in Section \ref{sec:gf}, the resulting metric reads as  (see Section X and Eqs.~(10.6) in~\cite{Thorne:1980ru})
\begin{subequations} \label{statmetric}
\begin{align}
g_{00} &= -1 + \frac{2 \mathbcal{I} }{r} + \frac{(0)\text{pole}}{r^2} + \sum_{l=2}^{\infty}\frac{1}{r^{l+1}} \left( \frac{2(2 l -1)!!}{l!} \mathbcal I_{A_l} N_{A_l} + (l-1)\text{pole} + \dots + (0)\text{pole} \right),\nn \\
g_{0j} &=  \sum_{l=1}^{\infty} \frac{1}{r^{l+1}} \left( -\frac{4l (2l-1)!!}{(l+1)!} \epsilon_{jpa_l} \mathbcal{S}_{pA_{l-1}} N_{A_{l}} + (l-1)\text{pole} + \dots + (0)\text{pole} \right),\label{met3}\\
g_{ij} &= \delta_{ij}\left(1+ \frac{2 \mathbcal{I} }{r}\right) + \frac{(0)\text{pole}}{r^2} + \sum_{l=2}^{\infty}\frac{1}{r^{l+1}} \left( \frac{2(2 l -1)!!}{l!} \mathbcal I_{A_l} N_{A_l} \delta_{ij} + (l-1)\text{pole} + \dots + (0)\text{pole} \right).\nn
\end{align}
\end{subequations}
Here $(0)\text{pole}$ is a constant monopole, $(1)\text{pole}$ a combination of $l=1$ spherical harmonics, etc. The tensors $N_{A_l}$ are defined in appendix \ref{sec:conv}. 
The coefficients $\mathbcal I_{A_l}$ and $\mathbcal S_{A_l}$ are defined as the mass multipole moments and the current multipole moments, respectively \cite{Thorne:1980ru}. 
In such canonical harmonic coordinates there is no mass dipole moment $\mathbcal{I}_i$. The mass and angular momentum are respectively $\mathbcal{I}$ and $\mathbcal{S_i}$. The STF version of the multipole moments $\mathbcal I_{A_l}$, $\mathbcal S_{A_l}$  can be translated in harmonic coefficients $I^{lm}$ and $ S^{lm}$ (see, \emph{e.g.}, \eqref{STF to harmonics}).

\subsection{Multipole charges}

The canonical charge of a solution $g_{\mu\nu}$ associated with the vector field $\xi^\mu$ is defined as a surface integral over a sphere and as an integral in phase space from the reference solution (here Minkowski $\eta_{\mu\nu}$) to the solution $g_{\mu\nu}$ \cite{Regge:1974zd,Iyer:1994ys,Barnich:2001jy,Barnich:2007bf} 
\begin{equation} \label{Fullcharge}
Q_{\xi}[g] = \frac{1}{8\pi G} \int_{\eta}^{g} \int_{S} \bold{k}_{\xi}[dg' ; g']. 
\end{equation}
The 2-form $\bold{k}_{\xi}$ is linear in its first argument and nonlinear in its second argument. It can be constructed from the Lagrangian, up to an ambiguous term proportional to $\mathcal L_\xi g_{\mu\nu} =  \nabla_{\mu}\xi_{\nu} +\nabla_{\nu}\xi_{\mu}$. There are two covariant prescriptions that exist in the literature to fix this ambiguity. We find convenient to label the surface charge by the parameter $\alpha$ multiplying the covariant ambiguous term. For $\alpha=0$, one has the Iyer-Wald charge \cite{Iyer:1994ys}. For $\alpha=1$, one has the Abbott-Deser \cite{Abbott:1981ff} or, equivalently, the Barnich-Brandt charge \cite{Barnich:2001jy}. The explicit expression for the surface charge in Einstein gravity is given in Appendix \ref{app:ch}.

Let us now evaluate the surface charge at spatial infinity for each of the three families of vector fields defined in \eqref{residual symm}. We first compute the spherical harmonic decomposition of each vector field. Let us discuss the $(l)$pole symmetries. We note that all subleading $(l+k)\text{pole}$ moments in \eqref{met3} ($k \geq 1$) do not contribute to the charges. This non-trivial property originates from the orthogonality of spherical harmonics, and the fact that the subleading $(l+k)\text{pole}$ moments are suppressed by negative powers of $r$ with respect to the leading $(l)\text{pole}$ moment of order $r^{-(l+1)}$. It can be checked by explicitly evaluating the charge. Since all non-linearities belong to such terms, only linear terms matter. The surface charge at spatial infinity can be equivalently defined using the linearized theory as
\begin{equation} \label{Fullchargelin}
Q_{\xi}[h] = \lim_{r \rightarrow \infty} \left( \frac{1}{8\pi G}  \int_{S} \bold{k}_{\xi}[h  ; \eta] \right),
\end{equation}
where $h_{\mu\nu}$ is the linear perturbation around $\eta_{\mu\nu}$ in the post-Minkowskian expansion of $g_{\mu\nu}$. The proof that all subleading monopole terms do not contribute to the definition of multipoles was essentially performed in Thorne \cite{Thorne:1980ru}. It also shows that any change of coordinates which preserves the canonical harmonic gauge does not change the definition of multipoles. 

With the help of a symbolic manipulation software, we obtain the following results: 
\paragraph{Mass multipole charges}
The mass multipole charges are associated to the multipole symmetries $K_{\epsilon}$. They read as
\begin{equation} \label{mass charge stationary}
Q_{K_{}}^{lm} =  \frac{2(2l+1)!!}{l!}  I^{lm}.
\end{equation}
Note that the result is independent of the $\alpha$ prescription. 

\paragraph{Current multipole charges}
The current multipoles charges are associated to the multipole symmetries $L_{\eps}$. They read as 
\begin{equation} \label{angular charge stationary}
Q_{L_{}}^{lm} =- \sqrt{\frac{l}{l+1}}\frac{(2 l-1)!!}{(l-1)!}\left(2 (l+2) +\alpha (l-1) \right) S^{lm}.
\end{equation}
The normalization constant depends upon the definition of the canonical charges through the $\alpha$ prescription. 
\paragraph{Momentum multipole charges}
The momentum multipole charges are identically zero for stationary configurations,
\bea
Q_{P_{}}^{lm} = 0. \label{Momentum multipoles charge}
\eea
There is no role to the momentum multipole charges in stationary Einstein gravity, but we expect that they might play a role in stationary configurations in other theories of gravity.

Equations \eqref{mass charge stationary}-\eqref{angular charge stationary}-\eqref{Momentum multipoles charge} are the main result of this section. Thorne's definition and the Noether charge definition agree up to a unique and well-defined normalization constant.

Let us make some additional comments. In the linear case, it turns out that the surface charge can be evaluated at any radius and yet gives the identical answers \eqref{mass charge stationary}-\eqref{angular charge stationary}-\eqref{Momentum multipoles charge}. This is due again to the orthogonality of spherical harmonics and the fact that, in the linear theory, each harmonic mode is associated with a particular radial falloff. The multipole symmetries generating such charges are therefore \emph{symplectic symmetries} in the linear theory in the terminology of \cite{Compere:2015bca,Compere:2015knw}.

A natural question is whether the charges associated with the multipole symmetries are affected by a change of gauge. In this paper, we did not investigate a generic change of gauge. Instead, we investigated the change of gauge from canonical harmonic gauge to generic mass-centered de Donder gauge in the linearized theory. The details are relegated to Appendix \ref{chg}. The result is that for linear stationary configurations, the current and momentum multipole charges $Q_{L_{\eps}}^{lm}$, $Q_{P_{\eps}}^{lm}$ are \emph{unaffected} by a change of gauge which preserves stationarity within de Donder gauge, while the mass multipole charges $Q_{K_{\epsilon}}^{lm}$ are affected, but only by gauge transformations of the form $\bx\to\bx+\nabla\eps$ with $\nabla^2 \eps = 0$, $\eps = O(r^{-1})$. 

The conclusion is that the current multipole moments can be computed in \emph{any} mass-centered de Donder gauge coordinate system. This is very useful in practice since work is only needed to reach de Donder gauge, but it is not necessary to further restrict to canonical harmonic gauge. However, the mass multipole moments need to be defined in the canonical harmonic gauge, or at least in mass-centered de Donder gauge where the gauge transformations $\bx\to \bx+\nabla\eps$ have been fixed according to the canonical prescription. This illustrates the gauge dependence in the definition of mass multipole moments.

\section{Multipole charges for linearized radiating solutions}
\label{sec:lin}

In this section, we turn our attention to dynamical radiating solutions in the linearized regime. We introduce the multipole charges at spatial infinity as conserved surface charges, while the source multipole moments are expressed in terms of surface charges in the near zone region. We will discuss the relation between these two and the implications of conservation laws for the time variation of the source multipole moments. We expect that the results for the linear theory can be extended to the non-linear theory essentially in the same way as discussed in the last section, but this problem is beyond the scope of this paper.

\subsection{Linearized metric in canonical harmonic gauge} \label{App:linearized metric}

Let $g_\mn=\eta_\mn+h_{\mu\nu}$ be the linearized metric describing the external gravitational field of an arbitrary isolated system with no incoming wave boundary conditions, and further excluding NUT or acceleration parameters which are considered unphysical. In the canonical harmonic gauge (also called $ACMC_\infty$ coordinates in~\cite{Thorne:1980ru}), $h_{\mu\nu}$ takes the following form
\begin{subequations} \label{perturbation metric}
\begin{align}
h_{00} &= \frac{2 \mathbcal{I} }{r} +\sum_{l=2}^{\infty} (-1)^{l} \frac{2}{l!} \partial_{A_{l}} \left(\frac{\mathbcal{I}_{A_{l}}(u)}{r}\right), \\
 h_{0j} &= -\sum_{l=1}^{\infty} (-1)^{l} \frac{4l}{(l+1)!} \partial_{q A_{l-1}}\left(\frac{\epsilon_{jpq} \mathbcal{S}_{pA_{l-1}}(u)}{r}\right) + \sum_{l=2}^{\infty} (-1)^{l} \frac{4}{l!} \partial_{A_{l-1}} \left(\frac{\dot{\mathbcal{I}}_{jA_{l-1}}(u)}{r}\right),\\
h_{ ij} &= h_{00} \delta_{ij} + \sum_{l=2}^{\infty} (-1)^{l} \left[\frac{4}{l!} \partial_{A_{l-2}} \left(\frac{\ddot{\mathbcal{I}}_{ijA_{l-2}}(u)}{r}\right) -  \frac{8l}{(l+1)!} \partial_{q A_{l-2}}\left(\frac{\epsilon_{pq(i} \dot{\mathbcal{S}}_{j)pA_{l-2}}(u)}{r}\right)\right].
\end{align}
\end{subequations}
The linearized metric is expressed in terms of symmetric trace-free (STF) tensors $\mathbcal{I}_{A_l}(u)$, $\mathbcal{S}_{A_l}(u)$ that are respectively the \textit{mass} and \textit{current} gravitational multipole moments. This is Thorne's definition in terms of metric components. The mass monopole $\mathbcal{I}$ is constant and is interpreted as the mass of the source. The current dipole $\mathbcal{S_i}$ is also constant and is interpreted as the angular momentum of the source. Because we are in the center of mass frame, no mass dipole moment $\mathbcal{I}_i$ is present.

The linearized perturbation $h_{\mu\nu}$ in Eqs.~\eqref{perturbation metric} can alternatively be expanded in terms of scalar, vector and tensor harmonics as follows
\begin{subequations} \label{non stat perturbation metric}
\begin{align}
 h_{00} &= \sum_{l=0}^{\infty} \sum_{m=-l}^{l}~\frac{2}{l!}~\sum_{k=0}^{l} \frac{1}{2^k k!}\frac{(l+k)!}{(l-k)!} \frac{~^{(l-k)}\!I^{lm}(u)Y^{lm}}{r^{k+1}}, \qquad \mbox{with} \; I^{1m} \equiv 0,\\
h_{0j} &= \sum_{l=1}^{\infty} \sum_{m=-l}^{l} \frac{4l}{(l+1)!} \sum_{k=0}^{l} \frac{1}{2^k k!}\frac{(l+k)!}{(l-k)!} \frac{~^{(l-k)}\!S^{lm}(u)~^{B}Y_{j}^{lm}}{r^{k+1}}  +\\
 & -\sum_{l=2}^{\infty} \sum_{m=-l}^{l}~\frac{4}{l!}~\sum_{k=0}^{l-1} \frac{1}{2^k k!}\frac{(l-1+k)!}{(l-1-k)!} \frac{~^{(l-k)}\!I^{lm}(u)}{r^{k+1}}  \left(^{E}Y_{j}^{lm} +\sqrt{\frac{l}{l+1}}~^{R}Y_{j}^{lm}\right) , \nonumber\\
 h_{ ij} &= h_{00} \delta_{ij} + \\
 &-\sum_{l=2}^{\infty} \sum_{m=-l}^{l}~\frac{8l}{(l+1)!}~\sum_{k=0}^{l-1} \frac{1}{2^k k!}\frac{(l-1+k)!}{(l-1-k)!} \frac{~^{(l-k)}\!S^{lm}(u)}{r^{k+1}}  \left(\sqrt{\frac{l-1}{l+2}} ~^{B1}T^{lm}_{ij} + ~^{B2}T^{lm}_{ij}\right) +\nonumber\\
 & + \sum_{l=2}^{\infty} \sum_{m=-l}^{l}~ \frac{4}{l!}~ \sum_{k=0}^{l-2} \frac{1}{2^k k!}\frac{(l-2+k)!}{(l-2-k)!}\frac{~^{(l-k)}\!I^{lm}(u)}{r^{k+1}} \times \nonumber\\
 &\quad\times \left(\sqrt{\frac{2(l-1)l}{(l+1)(l+2)}}~^{L0}T_{ij}^{lm} - \sqrt{\frac{(l-1)l}{(l+1)(l+2)}}~^{T0}T_{ij}^{lm} +2 \sqrt{\frac{l-1}{l+2}}~^{E1}T_{ij}^{lm} +~^{E2}T_{ij}^{lm} \right). \nonumber
\end{align}
\end{subequations}
The main elements of proof of the equivalence of STF tensor decomposition \eqref{perturbation metric} and the spherical harmonic decomposition \eqref{non stat perturbation metric} are presented in Appendix \ref{equiv}.

\subsection{Multipole charges}

The covariant phase space charges are defined in the linear theory as 
 \cite{Regge:1974zd,Iyer:1994ys,Barnich:2001jy,Barnich:2007bf} 
\begin{equation} \label{Fullcharge2}
Q_{\xi}[h] = \frac{1}{8\pi G}  \int_{S} \bold{k}_{\xi}[h ; \eta],
\end{equation}
where $\eta_{\mu\nu}$ is the Minkowski background and $h_{\mu\nu}$ the perturbation. The precise definition of the 2-form $\bold{k}_{\xi}$ is given in Appendix \ref{app:ch}. 

We now compute the Noether charges associated with the three sets of residual symmetries $L_{\epsilon}$, $K_{\epsilon}$ and $P_{\epsilon}$ presented in \eqref{residual symm}. The explicit values of the charges evaluated on a sphere at arbitrary retarded time $u$ and radius $r$ are given by the following expressions
\begin{subequations}\label{mainch}
\begin{align}
8\pi G ~Q^{lm}_{L_{}} &= \sum\limits_{p=0}^{l+1} ~ C_{L_{}}(p, l) ~r^{p} ~^{(p)}  S^{lm}(u) ,\label{mainch1}\\
8\pi G ~Q^{lm}_{K_{}} &= \sum\limits_{p=0}^{l+1} ~C_{K_{}}(p, l) ~r^{p} ~^{(p)} I^{lm}(u) +  8\pi G~ t \, Q^{lm}_{P_{}} \label{mainch2} ,\\
8\pi G ~Q^{lm}_{P_{}} &=\sum\limits_{p=0}^{l} ~C_{P_{}}(p, l) ~r^{p} ~^{(p+1)} I^{lm}(u) .\label{mainch3}
\end{align}
\end{subequations}
The coefficients $C_{L_{}}(p, l)$, $C_{K_{}}(p, l)$, $C_{P_{}}(p, l)$ are given in Appendix \ref{app:charges}. These charges are not well defined as such and require further insight in order to extract their physical content. We will discuss these charges in detail in the following. 

We will first define the \emph{conserved multipole charges} at spatial infinity. We then define the \emph{source multipole moments}. We finally define the \emph{radiation multipole moments} at null infinity as derived quantities. 

\subsubsection{Conserved multipole charges at spatial infinity}

Let us first assume stationarity at past of null infinity,
in the sense that  
\begin{subequations}\label{CK-condition}
    \begin{align}
     I^{lm}(u) = I^{lm} + \mathcal{O}\left(\dfrac{1}{u}\right), \qquad u\to -\infty ,\\
    S^{lm}(u) =  S^{lm} + \mathcal{O}\left(\dfrac{1}{u}\right), \qquad u\to -\infty .
    \end{align}
    \end{subequations}
We expect that we can relate these asymptotic conditions to the behavior of the Bondi news and Bondi mass \cite{Christodoulou:1993uv}.
These conditions imply that at spatial infinity (\emph{i.e.}, in the limit $r \rightarrow \infty$ at fixed $t=r + u$), 
\bea
\underset{\underset{u \rightarrow -\infty}{t \text{ fixed }}}{\lim} r^p\,{}^{(p)}  I^{lm}(u)  =0 =\underset{\underset{u \rightarrow -\infty}{t \text{ fixed }}}{\lim}  r^p \,{}^{(p)}  S^{lm}(u), \qquad \forall p \geq 1. \label{as}
\eea
This follows from $r^p\,{}^{(p)}  I^{lm}(u) \sim u^p \,{}^{(p)}I^{lm}(u)\sim \mathcal{O}(1/u)$ for $p\geq 1$. 

Under this asymptotic stationarity hypothesis, the finite stationary multipole charges \eqref{mass charge stationary}-\eqref{angular charge stationary}-\eqref{Momentum multipoles charge} are recovered at spatial infinity. These are the conserved multipole charges.

\subsubsection{Source multipole moments in the near-zone}
As we can see explicitly, the surface charges \eqref{mainch} are finite in the $r\to 0$ limit. This limit has a clear physical interpretation and gives the value of the surface charges in the near zone of the radiation zone. Indeed, one has in general that (see, \emph{e.g.}, Sec.~IX.D of \cite{Thorne:1980ru})
\begin{align}
\left|\dfrac{r^p \;{}^{(p)}  S^{lm}}{  S^{lm}}\right| \sim \left|\dfrac{r^p \;{}^{(p)}  I^{lm}}{  I^{lm}}\right|  \sim \left(\dfrac{r}{\lambda}\right)^p,
\end{align}
where $\lambda$ is the typical wavelength of the radiation generated by the sources. Accordingly, in the near zone where $r\ll \lambda$, we have
\begin{align}
Q=Q\Big\vert_{r=0} + \mathcal{O}\left(\dfrac{r}{\lambda}\right),
\end{align}
and all charges are finite. Moreover, all multipole moments are then functions of time $t$ since $u=t$ at $r=0$. In the post-Newtonian/post-Minkowskian matched asymptotic expansion scheme to solve Einstein's equations for a radiating non-linear system, there exist surface integrals defined in terms of the sources in the outer near-zone \cite{Blanchet:2004re}. Here, we will obtain the corresponding matching surface charges in the near-zone of the radiation region, in the linear approximation.  We will now explicitly describe which Noether charges give the source mass multipole moments $ I^{lm}(t)$ and current multipole moments $ S^{lm}(t)$.

\paragraph{Current multipole moments}

The current multipole moments are simply defined as the Noether charges associated with $L_\eps$ \eqref{residual symm} in the near zone. Their values are
\begin{align}
Q_L^{lm}\Big\vert_{r=0} = -\sqrt{\frac{l}{l+1}}\frac{(2l-1)!!}{( l-1)!} \left(2 (l+2) +\alpha (l-1) \right)  S^{lm}(t).
\end{align}
The numerical proportionality coefficient depends upon the detailed definition of the canonical charge through the $\alpha$ prescription.

\paragraph{Momentum multipole moments}
The momentum multipole moments are defined as the Noether charges associated with $P_\eps$ \eqref{residual symm} in the near zone. Their values are
\begin{align}
&Q_{P_{}}^{lm} \Big\vert_{r=0}  =-\frac{2 (2 l-3)!!}{l!} \left((2 l-1) l + (1-\delta _{1 l})\sqrt{\frac{l}{l+1}} \left[1 + (3-2l) l+\alpha  (l-1) (2 l+1)\right] \right)~^{(1)}I^{lm}(t).
\end{align}
Contrary to the stationary case, they are non-zero for radiating configurations. They are derived quantities in terms of the mass multipoles $I^{lm}$ and are therefore not fundamental, at least in Einstein gravity. They will however play a role in the definition of mass multipoles as we describe next.

\paragraph{Mass multipole moments}

Finally, the mass multipole moments are defined as the following combination of Noether charges in the near zone,
\begin{align}\label{sub}
 Q_{K_{}}^{lm}\Big\vert_{r=0}  - t\,  Q_{P_{}}^{lm} \Big\vert_{r=0}. 
\end{align}
After evaluation we find 
\begin{align}\label{sub2}
Q_{K_{}}^{lm}\Big\vert_{r=0}  - t\,  Q_{P_{}}^{lm} \Big\vert_{r=0} &= \frac{2(2l+1)!!}{l! }  I^{lm}(t),
\end{align}
which exactly reproduce Thorne's definition $I^{lm}(t)$, upon adjusting the normalization constant. Remark that both the Iyer-Wald and the Abbott-Deser-Barnich-Brandt definitions agree in this case since there is no $\alpha$ dependence in the numerical coefficient. 

The linear combination in the definition \eqref{sub} can be explained very naturally in two different ways. First, the definition \eqref{sub} is covariant under a time shift. The time dependence of the generator $K_{\epsilon}$ \eqref{residual symm} is exactly compensated by the subtracting term in \eqref{sub} and a time shift only shifts the source. More fundamentally, the definition \eqref{sub} generalizes to gravitating configurations and to generic mass multipole moments the definition of boost charge defined in standard field theories around Minkowski spacetime. For example, the Noether current of a scalar field of Lagrangian $\mathcal L=-\dfrac{1}{2}\p_\mu \phi \p^\mu \phi-V(\phi)$ associated with $x^\mu\to x^\mu-\xi^\mu$  is
\begin{align}
J^\mu_\xi&=\dfrac{\p \mathcal L}{\p (\p_\mu\phi)}\de_\xi\phi-\mathcal L\xi^\mu .
\end{align}
The boost charge along $x$ is given by 
\begin{align}\label{FT boost charge}
K_x=\int d^3x\, x\, \left(\dfrac{\dot{\phi}^2+|\nabla\phi|^2}{2}+V(\phi)\right)+t\int \dot{\phi}\nabla_x\phi=\int d^3x \,x\,\mathcal{H} +t\,P_x ~,
\end{align}
where $\mathcal H$ is the Hamiltonian density and $P_x$ is the total momentum along $x$. The $l = 1$ mass multipole $\int d^3x \,x\,\mathcal{H} $ therefore equates the boost charge minus $t$ times the momentum charge. Our definition of mass multipole moments \eqref{sub} is therefore totally natural in that respect. 

\subsubsection{Multipole charges at future null infinity}

Let us now describe how to define the multipole moments from Noether charges at future null infinity. We again assume the asymptotic stationarity hypothesis \eqref{CK-condition} at both  future and past of null infinity. 

For an arbitrary retarded time $u$, the charges \eqref{mainch} are formally infinite. However, they can be regularized by the finite part prescription of Blanchet-Damour \cite{1986RSPTA.320..379B}. After taking the finite part, we obtain in linearized Einstein gravity,
\bea
 ~\underset{\underset{r \rightarrow \infty}{u \text{ fixed}}}{\text{FP}} ~ \left( Q^{lm}_K - u Q^{lm}_P \right) &=& \frac{2(2l+1)!!}{l! }  I^{lm}(u) ,\label{Iu}\\
~\underset{\underset{r \rightarrow \infty}{u \text{ fixed}}}{\text{FP}} ~ Q^{lm}_L &=& -\sqrt{\frac{l}{l+1}}\frac{(2l-1)!!}{( l-1)!} \left(2 (l+2) +\alpha (l-1) \right) S^{lm}(u).
\eea
In other words, all terms proportional to $r$ are dropped and only the $r^0$ term is kept. This definition readily generalizes to an arbitrary theory of gravity. 

Let us now contrast these definitions with the standard \emph{radiative multipole moments}.
At linear order, one can switch to radiative (Bondi) coordinates $(U,R,\theta,\phi)$ by a simple change of coordinates, $U=u - \frac{2GM}{c^3}\log \left(r/r_0\right)$, $R=r$. The radiative mass and current multipole moments $U_{A_l}(U)$ and $V_{A_l}(U)$ are defined from the $1/R$ fall-off of the metric in radiative coordinates. It turns out that they match with the $l$-derivatives of the mass and current multipole moments, up to important non-linear corrections of $ \mathcal{O}(G)$ that encode tails, non-linear memory and further non-linear terms (see the review \cite{Blanchet:2013haa} and the latest update \cite{Marchand:2016vox})
\bea \label{rad}
U_{A_l}(U) &= &  ~^{(l)}\mathbcal I_{A_l}(U) + \mathcal{O}\left(\frac{G}{c^3}\right), \\
V_{A_l}(U) &= &  ~^{(l)}\mathbcal S_{A_l}(U)+ \mathcal{O}\left(\frac{G}{c^3}\right).
\eea
While the radiative multipole moments are proportional to the $l$-th derivative of the source multipole moments \eqref{rad}, we notice from Eq.~\eqref{Iu} that the Noether charges instead allow one to directly extract the source multipole moments without any derivatives close to null infinity. This is mainly because the multipole symmetries are proportional to $r^l$ and therefore extract information about the subleading $r^{-l}$ part of the metric. The multipole charges at null infinity are therefore more elementary than the radiative multipole moments.

\subsection{Conservation equation}

The multipole moments in linearized Einstein gravity can be simply read off from the components of the metric in canonical harmonic gauge. Now, it is advantageous already in Einstein gravity to reformulate these multipole moments in terms of Noether charges for the following reason. Canonical Noether charges obey a conservation law which allows us to relate the multipole charges of the system sourcing the linear solution at different times to the flux of multipole charges at null infinity. 

\begin{figure}[h]
    \captionsetup{width=0.8\textwidth}
    \centering
    \hspace{-4cm}
    \begin{subfigure}[c]{0.4\textwidth}
        \centering
        \includegraphics{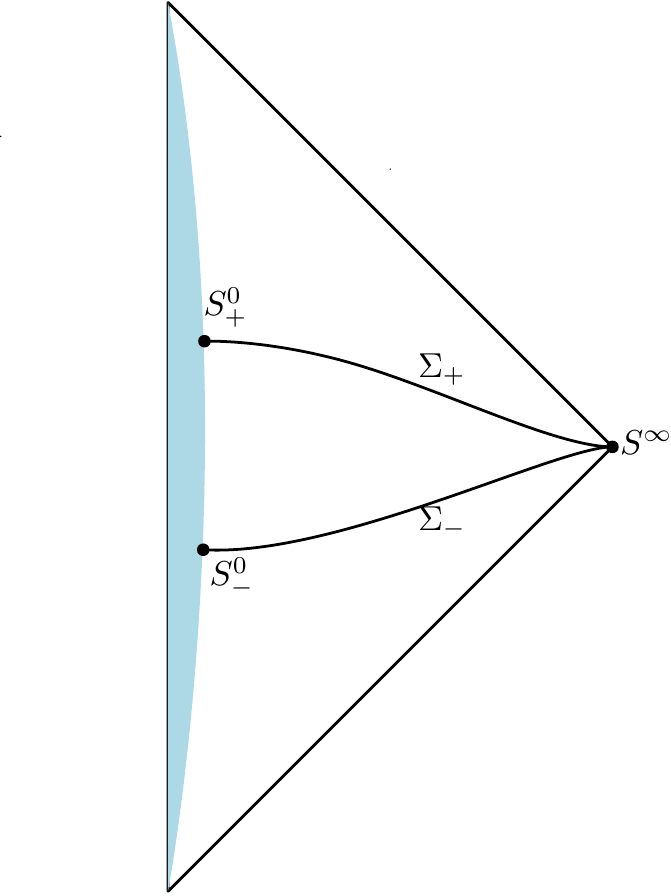}
    \end{subfigure}\hspace{-2cm}
    \begin{subfigure}[c]{0.4\textwidth}\centering\includegraphics{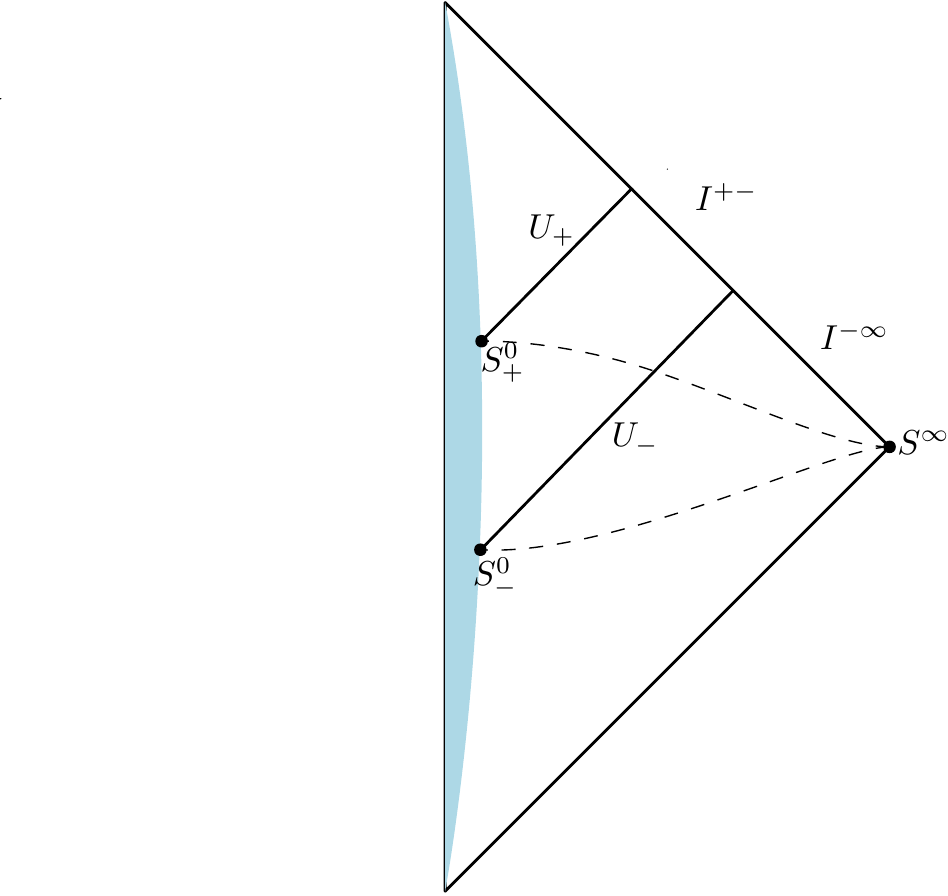}
    \end{subfigure}%
        \caption{(Left) Two constant time slices $\Sigma_{\pm}$ both asymptote to the sphere $S^\infty$ at spatial infinity. The coloured region $r \leq a$, where $a$ is the size of the source, is not described by the linear solution and contains the source. The inner boundaries are denoted as $S^0_\pm$. (Right) One can smoothly deform $\Sigma_{\pm}$ such that the difference of $ Q^{\text{rad}}_{\mathcal I^{+-}}$ between the two advanced times $u_+$ and $u_-$ represents the multipole moment flux through null infinity.}
    \label{fig:conservation}
\end{figure}

Let us derive this multipole moment conservation law. The generalized Noether theorem for gauge or diffeomorphism invariant theories  \cite{Wald:1993nt,Barnich:2000zw,Barnich:2001jy} implies that the surface charges obey the conservation equation
\bea
\int_{S^\infty_t} \boldsymbol k_\xi[h ;\eta] - \int_{S^0_t} \boldsymbol k_\xi[h ;\eta] = \int_{\Sigma_t} \boldsymbol \omega[h,\mathcal L_\xi \eta ; \eta],
\eea
where $\Sigma_t$ is a constant time hypersurface, whose boundary are two 2-spheres $S^\infty_t$ and  $S^0_t$. We choose $S^\infty_t$ to be the 2-sphere at spatial infinity and $S^0_t$ to be close to $r =0$, which is the near zone limit of the radiation zone. Let us choose two such hypersurfaces $\Sigma_{+}$ and $\Sigma_{-}$, respectively, at constant times $t^+$ and $t^-$, as shown in the left-hand-side of Figure \ref{fig:conservation}. Such constant time hypersurfaces are not boosted with respect to each other at spatial infinity and, in that sense, approach the same boundary sphere $S^\infty$. After fixing the ambiguity parameter $\alpha$ in the definition of the covariant charges, $\boldsymbol \omega$ is the Lee-Wald symplectic structure \cite{Lee:1990nz,Wald:1999wa} for $\alpha=0$, while $\boldsymbol \omega$ is the invariant symplectic structure \cite{Barnich:2007bf,Compere:2007az} for $\alpha=1$. 

By rearranging terms, one can write the above equation as 
\begin{align}
Q^{\text{source}}+Q^{\text{rad}}=Q^{\text{total}} \label{conser},
\end{align}
where
\begin{align}
Q^{\text{source}}= \int_{S^0_t} \boldsymbol k_\xi[h,\eta]\,,\qquad Q^{\text{total}} = \int_{S^\infty_t} \boldsymbol k_\xi[h,\eta] , \qquad  Q^{\text{rad}}= \int_{\Sigma_t} \boldsymbol \omega[h,\mathcal L_\xi \eta ; \eta]\,.
\end{align}
The source charge is the actual multipole charge of the source, as the near zone limit of the radiation zone matches with the far zone of the source zone; see, \emph{e.g.},~\cite{Blanchet:2013haa}. Under the hypothesis of asymptotic stationarity at initial retarded times, the multipole charges are time-independent at spatial infinity,
\begin{align}
\dfrac{d}{dt}Q^{\text{total}}=0.
\end{align}
The flux term is simply interpreted as the radiated multipole charge to future null infinity. More precisely, one can smoothly deform the hypersurfaces $\Sigma_\pm$ as shown in the right-hand-side of Figure  \ref{fig:conservation}, where we define $\Sigma_+ = U_+ \cup  I^{+-} \cup  I^{-\infty}$ and $\Sigma_- = U_- \cup  I^{-\infty}$. Here $U_\pm$ are null hypersurfaces at constant retarded time $u_\pm$ and $ I^{+-}$ is a surface at constant large $r$ close to null infinity.
We have the equality 
\bea\label{eq:6}
Q_{t^+}^{\text{source}} - Q_{t^-}^{\text{source}} = Q_{ I^{+-}  }^{\text{rad}} + Q_{U_+ }^{\text{rad}}  -  Q_{U_- }^{\text{rad}}.
\eea
The left-hand-side is the \emph{finite} difference between the source multipole charge at time $t^+$ and at time $t^-$. The individual terms on the right-hand-side are diverging but the difference is finite. It is useful to again define the finite part $\text{FP}$ of a diverging expression in $r$ as the $r^0$ term of the expression with $u$ fixed. In fact, we have 
\bea
~\underset{\underset{r \rightarrow \infty}{u \text{ fixed}}}{\text{FP}} ~ Q_{U_\pm}^{\text{rad}} = 0. 
\eea
The reason is that from the generalized Noether theorem, this expression is the difference of charge between $r \rightarrow \infty$ and $r=0$. From the explicit expression for the charges  (in the case of the mass multipole, a subtraction of two Noether charges is necessary, as \eqref{sub}), the only finite part is precisely the part at $r=0$ which therefore cancels out. The remaining finite part of the right-hand side of \eqref{eq:6} is the finite radiation flux in between $u^+$ and $u^-$, 
\bea
F^{+-} \equiv ~\underset{\underset{r \rightarrow \infty}{u \text{ fixed}}}{\text{FP}} ~ Q_{  I^{+-}  }^{\text{rad}} .
\eea
The multipole charge conservation law can be finally written as 
\bea
Q_{t^+}^{\text{source}} - Q_{t^-}^{\text{source}}  = F^{+-} . 
\eea
It follows from the generalized Noether theorem.

\section{Summary and further directions}
\label{sec:discussion}

We reformulated Thorne's definition of mass and current multipoles in Einstein gravity in terms of Noether charges associated with multipole symmetries, which are residual transformations of the harmonic gauge. Since Noether charges are defined for any theory of gravity, it allows to define in principle all mass and current multipoles for such an arbitrary theory. As an example, it is possible to determine from this method the multipole structure of the recently constructed black hole with scalar hair \cite{Herdeiro:2014goa} from the Noether charge formula of Einstein gravity coupled to a scalar field \cite{Barnich:2002pi} (see also Appendix \ref{app:ch}). 

We expressed  the multipole moments of General Relativity as conserved Noether charges at spatial infinity. We also derived the conservation laws of multipole charges under the hypothesis of no incoming radiation. With suitable junction conditions between the past of future null infinity and the future of past null infinity (which would generalize \cite{Strominger:2013jfa,Hawking:2016sgy}), one would relate the total change of source multipole moments between future and past timelike infinity to the corresponding fluxes at past and future null infinity. We expect that these conservation laws of canonical moments together with the conservation of canonical BMS supertranslation charges at spatial infinity \cite{Troessaert:2017jcm} (see also \cite{Compere:2011ve,Virmani:2011aa,Compere:2017knf})  underly the conservation laws of the Bondi mass and angular momentum aspects obtained in \cite{Hawking:2016sgy}.

We mainly discussed in this paper mass and current multipole moments. In a generic diffeomorphism invariant theory of gravity coupled to matter, there will be generically six independent STF tensors appearing in the linearized metric after gauge fixing; see Eq.~\eqref{genlin}.\footnote{Also, in a generic diffeomorphism invariant theory of gravity with lightcone propagating planar waves, there are six distinct polarization modes for the linearized perturbations \cite{Eardley:1973br}, simply because the $1/r$ behavior of the electric part of the Riemann tensor close to null infinity admits a decomposition into six distinct STF tensors. This leads to 6 distinct radiative multipole moments.} Therefore, we would need to define six independent sets of gravitational source multipole moments. In our approach, these would be built from six multipole Noether charges associated with multipole symmetries.

Starting from our ansatz \eqref{BC}, we identified four sets of multipole symmetries, which include the mass, current, and momentum multipole symmetries, but discarded the irreducible harmonic vector $\boldsymbol V$, since it does not play a role in Einstein gravity. We expect that it might play a role in more general theories. Note that $l = 0$ and $l =1$ harmonics of $\boldsymbol V$ are respectively the spatial scaling $x^i \p_i$ and the vector 
 $\xi^{(j)} =(x^i x^j - \frac{1}{3}x^2 \delta^{ij})\p_i$ which is not a standard symmetry. It is also straightforward to relax the ansatz \eqref{BC}, and consider as candidate multipole charges the set of eight charges 
\bea
& Q_{\eps \,\p_t+t\,\nabla \eps} - t \, Q_{\nabla \eps}, \quad 
Q_{\br\times\nabla\eps}, \quad
Q_{\nabla \eps},  \quad
Q_{V^i \p_i}, \\
& Q_{t \eps \,\p_t} - t \,Q_{\eps \,\p_t},  \quad
Q_{t \br\times\nabla\eps} - t \,Q_{\br\times\nabla\eps},  \quad
Q_{t \nabla \eps} - t \,Q_{\nabla \eps},  \quad
Q_{t V^i \p_i}-t \,Q_{V^i \p_i},
\eea
where $Q_\xi$ denotes the Noether charge associated with $\xi$. These linear combinations ensure covariance under explicit time shifts, as for the mass multipole moments \eqref{sub}. It is not clear to us, however, whether the six independent multipole moments can be identified from this set of eight combinations of Noether charges. 

Let us now discuss the connection of our results with gravitational electric-magnetic duality. In gauge theories, not every multipole moment might be associated with a Noether charge. Indeed, in the case of electromagnetism, electric multipole moments are Noether charges, but magnetic multipole moments cannot be expressed as Noether charges associated with the original gauge symmetry of Maxwell theory. In the vacuum, it exists a dual formulation in terms of a dual gauge field $A_D$ defined as $\star F = d A_D$. The emergent gauge symmetry of the dual formulation allows to define the magnetic multipoles of the original formulation in terms of Noether charges as a simple consequence of electric-magnetic duality. Contrary to electromagnetism, it turns out that in Einstein gravity all multipole charges \emph{are} Noether charges, as we have now demonstrated. In many instances the structure of dualities in Maxwell and in Einstein theories is similar, but not here. The reason of this asymmetry is that in Einstein gravity, the gauge parameter is a spacetime vector which itself admits duality transformations. It has been shown that, in the formulation of spatial infinity foliated by de Sitter slices of unit normal $n^\mu$, rotations and boosts are dual to each other, and the second subleading component of the electric part of the Weyl tensor is dual to its magnetic part \cite{Mann:2008ay,Compere:2011db}. This is the reason why the Lorentz charges can be expressed in two distinct but equivalent ways \cite{Ashtekar:1978zz,Mann:2006bd}. In fact, it is a matter of simple algebra to show that such a duality persists between the mass multipole symmetries and the current multipole symmetries: 
\bea
L_\eps = \frac{1}{2} \eps^{\mu\nu\rho\sigma }  n_\mu \p_\nu K_{\eps, \rho} \p_\sigma\, .
\eea
We therefore are led to conjecture that the Noether $l$-multipole charges can be expressed from the $(l+1)$--subleading component of either the electric or the magnetic part of Weyl tensor, as its $l=1$ Lorentz counterpart \cite{Compere:2011db}.

To conclude, the mass and current multipole symmetries are associated with an infinite number of conserved Noether charges. These outer symmetries provide a new concept to comprehend the asymptotic structure of gravity, or, after a suitable generalisation, of general gauge theories. 

\acknowledgments
The authors are grateful to D. Van den Bleeken for an early collaboration. G.C. and R.O. thank L. Blanchet for useful conversations. A.S. would like to thank useful discussions with B. Bonga, M. Campiglia and M. Geiller. The authors acknowledge the anonymous referee for useful comments that have improved the paper. G.C. is a Research Associate of the Fonds de la Recherche Scientifique F.R.S.-FNRS (Belgium). R.O. and G.C acknowledge the current support of the ERC Starting Grant 335146 ``HoloBHC". A.S. would like to thank the ERC Starting Grant 335146 funded visitor program at ULB and the support and hospitality of the Perimeter institute in last stages of this project. A.S. would like to thank the National Elites Foundation and Saramadan Club of Iran for their partial support. 

\newpage
\appendix

\section{Notation and conventions}\label{sec:conv}
Throughout this paper, we use the conventions in Thorne's paper \cite{Thorne:1980ru}. 
We adopt geometrized units, $G=1$, $c=1$, and Minkowski metric with mostly plus signature, $\eta = \mbox{diag}(-1, +1, +1, +1)$, to raise and lower spacetime indices. Spacetime indices are denoted by Greek letters, while spatial indices are denoted by Latin letters.
Multi-index tensors are abbreviated as
\begin{equation}
T_{A_{l}} \equiv T_{a_1a_2 \dots a_l}.
\end{equation}
Round brackets stand for symmetrization $T_{(ab)}=\frac{1}{2}(T_{ab}+T_{ba})$, while squared brackets stand for anti-symmetrization $T_{[ab]}=\frac{1}{2}(T_{ab}-T_{ba})$. The unit radial vector in the $x_{i}$ direction is denoted as $n_{i} = x_{i}/r$, where $r = \vert \bold{x} \vert = \sqrt{x^2+y^2+z^2}$. The transverse projection tensor $P_{ij} = \delta_{ij} - n_{i}n_{j}$ is used to construct the \emph{transverse} (T) part of a tensor field
\begin{equation}
[T_{a_1 a_2 \dots a_l}]^{T} = P_{a_1 a'_1}P_{a_2 a'_2} \dots P_{a_l a'_l}T_{a'_1 a'_2 \dots a'_l}.
\end{equation}
The \emph{transverse-traceless} (TT) part of a rank-two tensor field is defined as 
\begin{equation}
[T_{ij}]^{TT} = P_{ia}P_{jb}T_{ab} - \frac{1}{2}P_{ij} \left(P_{ab}T_{ba}\right).
\end{equation}
The \emph{symmetric-transverse-traceless} (STT) part of a tensor field of rank-two $[T_{ij}]^{STT}$ is the symmetric part of $[T_{ij}]^{TT}$. Finally, we denote by capital script letters those tensor fields that are fully \emph{symmetric and trace-free} (STF):
\begin{equation}
\begin{gathered}
\mathbcal{T}_{A_l} \equiv [T_{A_{l}}]^{STF} = \sum_{n=0}^{[l/2]}a_{n} \delta_{(a_1 a_2} \dots \delta_{a_{2n-1}a_{2n}} S _{a_{2n+1} \dots a_{l}) j_1 j_1 \dots j_n j_n},\\
a_{n} = (-1)^n \frac{l!(2l-2n-1)!!}{(l-2n)!(2l-1)!!(2n)!!},
\end{gathered}
\end{equation}
where $S_{A_l} = [T_{A_l}]^{S}$ is the fully symmetric tensor field constructed from $T_{A_l}$. As an explicit example of a tensor of rank-2, one has $\mathbcal{T}_{ij} = [T_{ij}]^S - \frac{1}{3}\delta_{ij} T_{kk}$.

\section{Spherical harmonics}\label{sec-spherical harmonics}
In this Appendix, we recall the definitions and the main properties of spherical harmonics used throughout this paper and strictly necessary for our computation. We refer the reader to \cite{Thorne:1980ru} for further properties.

\subsection{Scalar spherical harmonics}
The \emph{pure-orbital scalar spherical harmonics} $Y^{lm}(\theta, \phi)$ are eigenfunctions of the squared \emph{orbital angular momentum} operator $\bold{L}^2$, which is defined as the angular part of the Laplacian operator $\nabla^2$ in spherical coordinates $(r, \theta, \phi)$:
\begin{equation}
\nabla^2  = \frac{1}{r^2}\frac{\partial}{\partial r}\left( r^2 \frac{\partial}{\partial r} \right) - \frac{\bold{L}^2}{r^2}, \qquad \bold{L}^2 = - \left[\frac{1}{\sin\theta}\frac{\partial}{\partial \theta} \left(\sin\theta\frac{\partial}{\partial \theta} \right) + \frac{1}{\sin^2 \theta}\frac{\partial^2}{\partial \phi^2}\right].
\end{equation}
The minus sign in the definition of $\bold{L}^2$ is taken in order to have positive eigenvalues $l(l+1)$. Explicitly, the pure-orbital scalar spherical harmonics are given by
\begin{equation}
Y^{lm}(\theta, \phi) = C^{lm} e^{i m \phi} P^{lm}(\cos(\theta)),
\end{equation}
where $P^{lm}(x)$ are the associated Legendre polynomials 
\bea
P^{lm}(x) = (-1)^{m} \left(1-x^2\right)^{m/2}\frac{d^m P^{l}(x)}{dx^m},
\eea
and $C^{lm}$ are the normalization factors 
\begin{equation}
C^{lm} = (-1)^{m} \sqrt{\frac{2l+1}{4\pi} \frac{(l-m)!}{(l+m)!}}.
\end{equation} 
Scalar spherical harmonics are orthonormal on the two-sphere $S^2$
\begin{equation}
\int Y^{lm} \bar{Y}^{l' m'} d\Omega = \delta_{l l'} \delta_{m m'},
\end{equation}
where $\bar{Y}^{l m} = (-1)^m Y^{l\, -m}$ is the complex conjugate of $Y^{lm}$ and $d\Omega=\sin\theta d\theta d\phi$.

Any smooth scalar function $f(\theta, \phi)$ can be expanded either in pure-orbital scalar spherical harmonics with complex coefficients or in terms of tensor product of $l$ unit radial vectors with STF-$l$ tensor coefficients:
\begin{equation} \label{SHtoSTF}
f(\theta, \phi) = \sum_{l=0}^{\infty}\sum_{m=-l}^{m=l} F^{lm} Y^{lm} = \sum_{l=0}^{\infty} \mathbcal{F}_{A_{l}} N_{A_l}.
\end{equation}
In the main text we sometimes switch between $Y^{lm}$ and $Y_{lm}$ for the ease of notation.

For practical applications, it is convenient to expand functions in terms of the \textit{real} spherical harmonics $Y_l^m$ which we define in terms of standard complex spherical harmonics $Y_{l m}$ as 
\begin{align}\label{realY}
C^{lm}Y_l^m&\equiv\left\lbrace\begin{array}{lc}
\frac{i}{\sqrt{2}}(Y_{l, m}-(-1)^m Y_{l ,-m})&m<0,\\
Y_{l, 0}&m=0,\\
\frac{1}{\sqrt{2}}(Y_{l, -m}+(-1)^m Y_{l, m})&m>0.\\
\end{array}\right.
\end{align}
Note that we have removed the normalization factor for the real harmonics. For example, the first few real harmonics will be 
\begin{align}\label{realY1}
Y_0^0=1,\qquad Y_1^1=\dfrac{x}{r},\qquad Y_1^{-1}=\dfrac{y}{r},\qquad Y_1^0=\dfrac{z}{r}.
\end{align}

\subsection{Vector spherical harmonics}\label{sec-VSH}
The \emph{pure-spin vector spherical harmonics} of magnetic ($B-$), electric ($E-$) and radial ($R-$) type are defined as \cite{Thorne:1980ru} 
\begin{subequations} \label{pure-spin VSH}
\begin{align}
~^{B}\bold{Y}^{lm} &= \frac{1}{\sqrt{l(l+1)}} i \bold{L}Y^{lm} = \bold{n} \times {}^{E}\bold{Y}^{lm} , \label{VSH-B STF def}\\
~^{E}\bold{Y}^{lm} &=\frac{1}{\sqrt{l(l+1)}} r \nabla Y^{lm}= -\bold{n} \times {}^{B}\bold{Y}^{lm} ,\\
~^{R}\bold{Y}^{ lm} &= \bold{n}\, Y^{lm}.
\end{align}
\end{subequations}
Here, $\nabla $ is the Euclidean gradient operator, $\bold{L} = -i \bold{r} \times \nabla$ is the orbital angular momentum operator, and $\bold{n}$ is the unit radial vector. The B- and E-type vector spherical harmonics are defined for $l \geq 1$ and are identically zero for $l=0$. All of them are orthonormal, \emph{i.e.},
\begin{equation}
\int ~^{J}\bold{Y}^{lm} \cdot ~^{J'}\bar{\bold{Y}}^{l' m'} d\Omega = \delta_{JJ'}\delta_{l l'}\delta_{m m'},\qquad \forall J=B,\,E,\,R
\end{equation}
where $~^{J}\bar{\bold{Y}}^{l m} = (-1)^m ~^{J}\bold{Y}^{l\;-m}$ is the complex conjugated.\\
The B- and E-type are transverse, while the R-type is radial:
\begin{equation}
\bold{n}\cdot~^{B}\bold{Y}^{lm} =0, \quad \bold{n}\cdot~^{E}\bold{Y}^{lm} =0, \quad \bold{n}\cdot~^{R}\bold{Y}^{lm} = Y^{lm}.
\end{equation}
A useful property of the pure-spin vector spherical harmonics is their divergence
\begin{equation}
\nabla \cdot~^{B}\bold{Y}^{lm}  = 0, \quad
\nabla \cdot~^{E}\bold{Y}^{lm} = - \sqrt{l(l+1)} \frac{Y^{lm}}{r}, \quad   \nabla \cdot~^{R}\bold{Y}^{lm} = 2  \frac{Y^{lm}}{r}.
\end{equation}
The magnetic type pure-spin vector spherical harmonics are eigenvectors of $\bold{L}^2$, while the electric and radial type are not. Explicit computations give  
\begin{subequations} \label{LL on pure-spin}
\begin{align}
\bold{L}^2 ~^{B}\bold{Y}^{lm} &= l(l+1)~^{B}\bold{Y}^{lm}, \\
\bold{L}^2 ~^{E}\bold{Y}^{lm} &= l(l+1)~^{E}\bold{Y}^{lm} -2 \sqrt{l(l+1)}~^{R}\bold{Y}^{lm},\\
\bold{L}^2 ~^{R}\bold{Y}^{lm} &=(l(l+1)+2)~^{R}\bold{Y}^{lm} -2 \sqrt{l(l+1)}~^{E}\bold{Y}^{lm}.
\end{align}
\end{subequations}
The STF version of the pure-spin vector spherical harmonics are obtained by inserting the decomposition of the scalar spherical harmonics $Y^{lm}$ in terms of STF-$l$ tensors $ \mathbcal{Y}^{lm}_{A_{l}}$ (defined in Eq.~(2.12) of \cite{Thorne:1980ru}), $Y^{lm} = \mathbcal{Y}^{lm}_{A_{l}} N_{A_{l}}$, in the defining equations~\eqref{pure-spin VSH}. The result is given by \cite{Thorne:1980ru}   
\begin{subequations} \label{pure-spin vector}
\begin{align}
~^{B}Y^{lm}_{i} &=\sqrt{\frac{l}{l+1}}\epsilon_{i p q} n_p \mathbcal{Y}^{lm}_{qA_{l-1}} N_{A_{l-1}} , \label{VSH-B STF}\\
~^{E}Y^{lm}_{i} &= \sqrt{\frac{l}{l+1}} \left[\mathbcal{Y}^{lm}_{i A_{l-1}} N_{A_{l-1}}\right]^{T},\\
~^{R}Y^{lm}_{i} &= n_{i}\mathbcal{Y}^{lm}_{A_{l}} N_{A_{l}}.
\end{align}
\end{subequations}
For completeness, we also write the STF version of the \emph{pure-orbital vector spherical harmonics} that are eigenvectors of $\bold{L}^2$ with eigenvalues $l'(l'+1)$ and $l'=(l-1, l, l+1)$:
\begin{subequations} \label{pure-orbital vector}
\begin{align}
Y_{i}^{l-1, lm} &= \sqrt{\frac{l}{2l+1}} \mathbcal{Y}^{lm}_{i A_{l-1}}N_{A_{l-1}},\\
Y_{i}^{l, lm} &= -i \sqrt{\frac{l}{l+1}} \epsilon_{i p q} n_p \mathbcal{Y}^{lm}_{qA_{l-1}} N_{A_{l-1}},\\
Y_{i}^{l+1, lm} &= -\sqrt{\frac{2l+1}{l+1}}\left[n_{i}\mathbcal{Y}^{lm}_{ A_{l}}N_{A_{l}}-\left(\frac{l}{2l+1}\right) \mathbcal{Y}^{lm}_{i A_{l-1}}N_{A_{l-1}} \right].
\end{align}
\end{subequations}
Any spatial vector field $v_{i}$ can be expanded either in pure-spin vector harmonics or in pure-orbital vector harmonics:
\begin{subequations}
\begin{align}
v_{i} &=   \sum_{l=0}^{\infty}\sum_{m=-l}^l  R^{lm} ~^{R}Y^{lm}_{i}  +  \sum_{l=1}^{\infty}\sum_{m=-l}^l \left( B^{lm} ~^{B}Y^{lm}_{i} + E^{lm} ~^{E}Y^{lm}_{i} \right),\\
&=\sum_{l=0}^{\infty} n_i \mathbcal{R}_{A_l} N_{A_l}  +  \sum_{l=1}^{\infty}\left( \epsilon_{i p q} n_p \mathbcal{B}_{qA_{l-1}} N_{A_{l-1}} + \left[\mathbcal{E}_{i A_{l-1}}N_{A_{l-1}} \right]^T \right),
\end{align}
\end{subequations}
from which one deduces, by using the STF decomposition of the pure-spin vector harmonics in Eqs.~\eqref{pure-spin vector}, the relations between the STF coefficients and the vector harmonic coefficients \cite{Thorne:1980ru}
\begin{subequations}
\begin{align}
\mathbcal{B}_{A_{l}} &= \sqrt{\frac{l}{l+1}}\sum_{m=-l}^{l} B^{lm} \mathbcal{Y}^{lm}_{A_{l}},\\
\mathbcal{E}_{A_{l}} &= \sqrt{\frac{l}{l+1}}\sum_{m=-l}^{l} E^{lm} \mathbcal{Y}^{lm}_{A_{l}},\\
\mathbcal{R}_{A_{l}} &=\sum_{m=-l}^{l} R^{lm} \mathbcal{Y}^{lm}_{A_{l}}.
\end{align}
\end{subequations}

\subsection{Tensor spherical harmonics} 
The \emph{pure-spin tensor spherical harmonics} are defined as \cite{Thorne:1980ru} 
\begin{subequations}
\begin{align}
~^{L0}\bold{T}^{lm} &= \bold{n} \otimes \bold{n} Y^{lm},\\
~^{T0}\bold{T}^{lm} &= \frac{1}{\sqrt{2}} \bold{P}Y^{lm},\\
~^{E1}\bold{T}^{lm} &= \sqrt{\frac{2}{l(l+1)}} r \left[ \bold{n} \otimes \nabla Y^{lm}\right]^{S},\\
~^{B1}\bold{T}^{lm} &= \sqrt{\frac{2}{l(l+1)}} \left[ \bold{n} \otimes i \bold{L} Y^{lm}\right]^{S},\\
~^{E2}\bold{T}^{lm} &= \sqrt{2\frac{(l-2)!}{(l+2)!}} r^2 \left[ \nabla \nabla Y^{lm}\right]^{STT},\\
~^{B2}\bold{T}^{lm} & = \sqrt{2\frac{(l-2)!}{(l+2)!}} i r \left[ \nabla \bold{L} Y^{lm}\right]^{STT}.
\end{align}
\end{subequations}
The longitudinal (L0) and transverse (T0) spin-0 tensor harmonics are defined for $l \geq 0$, the E1- and B1-type spin-1 tensor harmonics for $l \geq 1$, and the transverse and traceless (E2 and B2) spin-2 tensor harmonics for $l \geq 2$. All of them are orthonormal, in the sense that
\begin{equation}
\int \mbox{Tr} \left(~^{JS}\bold{T}^{lm} ~^{J'S'}\bar{\bold{T}}^{l'm'}\right) d\Omega = \delta_{JJ'} \delta_{S S'} \delta_{l l'}\delta_{m m'},
\end{equation}
where  $~^{JK}\bar{\bold{T}}^{lm} = (-1)^m ~^{JK}\bold{T}^{l \; -m} $ is the complex conjugated. Pure-spin tensor harmonics obey the following directional properties 
\begin{subequations}
\begin{align}
\bold{n} \cdot ~^{L0}\bold{T}^{lm} &=  ~^{R}\bold{Y}^{lm}, \qquad 
\bold{n} \cdot ~^{T0}\bold{T}^{lm} = 0, \qquad
\bold{n} \cdot ~^{E1}\bold{T}^{lm} = \frac{1}{\sqrt{2}}~^{E}\bold{Y}^{lm}, \\
\bold{n} \cdot ~^{B1}\bold{T}^{lm} &= \frac{1}{\sqrt{2}} ~^{B}\bold{Y}^{lm},\qquad
\bold{n} \cdot ~^{E2}\bold{T}^{lm} =0, \qquad
\bold{n} \cdot ~^{B2}\bold{T}^{lm} = 0.
\end{align}
\end{subequations}
The trace of the pure-spin tensor harmonics read as
\begin{subequations}
\begin{align}
\mbox{Tr}\left(^{L0}\bold{T}^{lm}\right) &= Y^{lm}, \\
\mbox{Tr}\left(^{T0}\bold{T}^{lm}\right) &= \sqrt{2}Y^{lm}, \\
\mbox{Tr}\left(^{JS}\bold{T}^{lm}\right) & =0, \qquad \hspace{5mm} JS=(E1,E2,B1,B2)
\end{align}
\end{subequations}
and their divergence is given by 
\begin{subequations}
\begin{align}
\nabla \cdot  ~^{L0}\bold{T}^{lm} &=  2 \frac{~^{R}\bold{Y}^{lm}}{r}, \\ 
\nabla \cdot  ~^{T0}\bold{T}^{lm} &= \frac{1}{\sqrt{2}}\left[\sqrt{l(l+1)}\frac{~^{E}\bold{Y}^{lm}}{r} -2 \frac{~^{R}\bold{Y}^{lm}}{r}\right], \\
\nabla \cdot  ~^{E1}\bold{T}^{lm} &= \frac{1}{\sqrt{2}}\left[3 \frac{~^{E}\bold{Y}^{lm}}{r} - \sqrt{l(l+1)} \frac{~^{R}\bold{Y}^{lm}}{r}\right], \\
\nabla \cdot  ~^{B1}\bold{T}^{lm} &= \frac{3}{\sqrt{2}} \frac{~^{B}\bold{Y}^{lm}}{r},\\
\nabla \cdot  ~^{E2}\bold{T}^{lm} &=- \sqrt{\frac{(l+2)!}{2(l-2)!}\frac{1}{l(l+1)}}\frac{~^{E}\bold{Y}^{lm}}{r} , \\
\nabla \cdot  ~^{B2}\bold{T}^{lm} &= - \sqrt{\frac{(l+2)!}{2(l-2)!}\frac{1}{l(l+1)}}\frac{~^{B}\bold{Y}^{lm}}{r}.
\end{align}
\end{subequations}

The STF version of the pure-spin tensor spherical harmonics are derived in Eqs.~(2.39) of \cite{Thorne:1980ru}
\begin{subequations} \label{pure-spin tensor}
\begin{align}
~^{L0}T_{ij}^{lm} &=  n_i n_j \mathbcal{Y}^{lm}_{A_l}N_{A_l}, \\ 
~^{T0}T_{ij}^{lm} &= \frac{1}{\sqrt{2}}P_{ij}\mathbcal{Y}^{lm}_{A_l}N_{A_l} , \\
~^{E1}T_{ij}^{lm} &= \sqrt{\frac{2l}{l+1}} \left(n_{(i} \mathbcal{Y}^{lm}_{j)A_{l-1}} N_{A_{l-1}} - n_i n_j \mathbcal{Y}^{lm}_{A_l} N_{A_l} \right), \\
~^{B1}T_{ij}^{lm} &= \sqrt{\frac{2l}{l+1}} n_{(i} \epsilon_{j)pq}n_p \mathbcal{Y}^{lm}_{q A_{l-1}}N_{A_{l-1}},\\
~^{E2}T_{ij}^{lm} &= \sqrt{\frac{2(l-1)l}{(l+1)(l+2)}}\left[\mathbcal{Y}^{lm}_{i j A_{l-2}}N_{A_{l-2}} \right]^{TT} , \\
~^{B2}T_{ij}^{lm} &= \sqrt{\frac{2(l-1)l}{(l+1)(l+2)}} \left[n_p \epsilon_{pq (i}\mathbcal{Y}^{lm}_{j) q A_{l-2}}N_{A_{l-2}} \right]^{TT}.
\end{align}
\end{subequations}
It is useful to notice that the pure-spin tensor spherical harmonics of type E2 and B2 are proportional to the TT part of the following pure-orbital tensor spherical harmonics \cite{Thorne:1980ru}
\begin{subequations} \label{pure-orbital tensor}
\begin{align}
T_{ij}^{2 \; l-2, lm} &= \sqrt{\frac{(l-1)l}{(2l-1)(2l+1)}}\mathbcal{Y}^{lm}_{i j A_{l-2}}N_{A_{l-2}},\\
T_{ij}^{2 \; l-1, lm} &= i \sqrt{\frac{2(l-1)l}{(l+1)(2l+1)}}n_p \epsilon_{pq (i}\mathbcal{Y}^{lm}_{j) q A_{l-2}}N_{A_{l-2}},
\end{align}
\end{subequations}
that are eigenfunctions of the squared angular momentum operator $\bold{L}^2$ with eigenvalues $l'(l' +1)$ with $l' = l-2$ for $T_{ij}^{2 \; l-2, lm}$ and $l' =l-1$ for $T_{ij}^{2 \; l-1, lm}$. These pure-orbital tensor harmonics can be rewritten as linear combination of pure-spin tensor harmonics as follows \cite{Thorne:1980ru}:
\begin{subequations} \label{pure-O to pure-S}
\begin{align}
T_{ij}^{2 \; l-2, lm} &= \sqrt{\frac{(l-1)l}{(2l-1)(2l+1)}}~^{L0}T_{ij}^{lm} - \sqrt{\frac{(l-1)l}{2(2l-1)(2l+1)}}~^{T0}T_{ij}^{lm}  \nonumber\\
&\quad +\sqrt{\frac{2(l-1)(l+1)}{(2l-1)(2l+1)}}~^{E1}T_{ij}^{lm} +\sqrt{\frac{(l+1)(l+2)}{2(2l-1)(2l+1)}}~^{E2}T_{ij}^{lm}, \\
T_{ij}^{2 \; l-1, lm} &= i \sqrt{\frac{l-1}{2l+1}}~^{B1}T_{ij}^{lm} +  i \sqrt{\frac{l+2}{2l+1}}~^{B2}T_{ij}^{lm}.
\end{align}
\end{subequations}

\section{Proofs}

\subsection{Equivalence between Eqs.~\eqref{perturbation metric} and \eqref{non stat perturbation metric}}
\label{equiv}

The multi-index derivative of a STF-$l$ tensor, which is function of the retarded time $u=t-r$, can be expanded in terms of its derivatives as \cite{Thorne:1980ru}
\begin{equation} \label{der1}
\partial_{A_{l}} \left(\frac{\mathbcal{A}_{A_{l}}(u)}{r}\right) = (-1)^l \sum_{k=0}^{l}c_{kl}\frac{~^{(l-k)}\mathbcal{A}_{A_{l}}(u) N_{A_{l}}}{r^{k+1}}, \qquad c_{kl} = \frac{1}{2^k k!}\frac{(l+k)!}{(l-k)!}
\end{equation}
where $^{(l-k)}\mathbcal{A}_{A_{l}}(u)$ is the $(l-k)$-th derivative of $\mathbcal{A}_{A_{l}}(u)$ with respect to the retarded time coordinate $u$. In particular, we have 
\bea
\partial_{A_{l}} \left(\frac{1}{r}\right) = \frac{(-1)^l (2l - 1)!!}{r^{l+1}}N_{A_l}. \label{rel1r}
\eea
The relations between the STF-$l$ coefficients (and their derivatives) and the harmonic coefficients can be read from the STF-$l$ version of the scalar, vector and tensor harmonics in Eqs.~\eqref{SHtoSTF}-\eqref{pure-spin vector}-\eqref{pure-spin tensor}:
\begin{subequations} \label{STF to harmonics}
\begin{align}
~^{(l-k)}\!\mathbcal{I}_{A_{l}} &= \sum_{m=-l}^{l} ~^{(l-k)}\!I^{lm}\mathbcal{Y}^{lm}_{A_l},\\
~^{(l-k)}\!\mathbcal{I}_{jA_{l-1}} &= \sqrt{\frac{l}{l+1}}\sum_{m=-l}^{l} ~^{(l-k)}\!I^{lm}\mathbcal{Y}^{lm}_{j A_{l-1}},\\
~^{(l-k)}\!\mathbcal{S}_{jA_{l-1}} &= \sqrt{\frac{l}{l+1}}\sum_{m=-l}^{l} ~^{(l-k)}\!S^{lm}\mathbcal{Y}^{lm}_{j A_{l-1}},\\
~^{(l-k)}\!\mathbcal{I}_{ij A_{l-2}} &=  \sqrt{\frac{2(l-1) l}{(l+1)(l+2)}} \sum_{m=-l}^{l} ~^{(l-k)}\!I^{lm}\mathbcal{Y}^{lm}_{ij A_{l-2}},\\
~^{(l-k)}\!\mathbcal{S}_{ij A_{l-2}} &= \sqrt{\frac{2(l-1) l}{(l+1)(l+2)}} \sum_{m=-l}^{l} ~^{(l-k)}\!S^{lm}\mathbcal{Y}^{lm}_{ij A_{l-2}}.
\end{align}
\end{subequations}

The proof then uses the definitions of the \emph{pure-orbital vector harmonics} $(\bold{Y}^{l-1, lm}, \bold{Y}^{l, lm}, \bold{Y}^{l+1, lm})$ in Eqs.~\eqref{pure-orbital vector} and their relations with the \emph{pure-spin vector harmonics} $(~^{B}\bold{Y}^{lm}, ~^{E}\bold{Y}^{lm}, ~^{R}\bold{Y}^{lm})$ in Eqs.~\eqref{pure-spin VSH}. Along the same lines, we use the definitions of the \emph{pure-orbital tensor harmonics} $(\bold{T}^{2\; l-1, lm}, \bold{T}^{2\; l-2, lm})$ in Eqs.~\eqref{pure-orbital tensor} and their relations with the expressions of the \emph{pure-spin vector harmonics} $(^{L0}\bold{T}^{lm}, ~^{T0}\bold{T}^{lm}, ~^{B1}\bold{T}^{lm}, ~^{E1}\bold{T}^{lm}, ~^{B2}\bold{T}^{lm}, ~^{E2}\bold{T}^{lm})$ in Eqs.~\eqref{pure-O to pure-S}.

After tedious but straightforward algebra, we get the harmonic decomposition in Eq.~\eqref{non stat perturbation metric}.
Notice that the harmonic decomposition contains all the the degrees of freedom of a spin-two symmetric tensor field, namely the pure-longitudinal and the pure-transverse spin-zero modes $(^{L0}\bold{T}^{lm}, ~^{T0}\bold{T}^{lm})$, the mixed transverse and longitudinal spin-one modes $(^{B1}\bold{T}^{lm}, ~^{E1}\bold{T}^{lm})$, and the physical transverse and traceless modes $(^{B2}\bold{T}^{lm}, ~^{E2}\bold{T}^{lm})$.

\subsection{General residual transformation}
\label{genres}

We want to find the most general class of solutions to $\square_{\eta} \xi^{\mu}=0$ in terms of scalar and vector harmonics. We shall not use a Fourier transformation since linear modes in $t$ are important solutions. Let us start with the most general vector field $\xi$ decomposed in terms of scalar spherical harmonics $Y^{lm}$ and pure-spin vector spherical harmonics $( ^{B}\bold{Y}^{l m}, ~^{E}\bold{Y}^{l m}, ~^{R}\bold{Y}^{l m})$,
\begin{equation} \label{vector field}
\begin{gathered}
\xi(t, r, \theta, \phi) = \sum_{l=0}^{\infty} \sum_{m=-l}^{l} \left( S_{l m}(t, r) ~Y^{l m}(\theta,\phi) \partial_{0} + R_{l m}(t, r) ~^{R}Y_{i}^{l m}(\theta,\phi) \partial_{i} \right) +\\
 +\sum_{l=1}^{\infty} \sum_{m=-l}^{l} \left(B_{l m}(t, r) ~^{B}Y_{i}^{l m}(\theta,\phi) + E_{l m}(t, r) ~^{E}Y_{i}^{l m}(\theta,\phi) \right)\partial_{i},
 \end{gathered}
\end{equation}
where the angular dependence is encoded in the spherical harmonics and the time and radial dependence is encoded in the coefficients of the linear combination. 

For definiteness, let us study the solutions to $\square_{\eta} \xi^{0}=0$ in full detail. The harmonic gauge condition amounts to the following partial differential equation for the coefficients $S_{lm}(t, r)$
\begin{equation} \label{S constraint}
r^2\left(-\partial^2_{t}S_{lm} + \partial^2_{r}S_{lm}\right) + 2 r \partial_{r}S_{lm} - l(l+1)S_{lm}=0,
\end{equation}
where we used the property that $\bold{L}^2 Y^{lm}= l(l+1)Y^{lm}$. Since $\p_t$ is a Killing vector, we can assume the separation of variables $S_{lm}(t,r) =f_{lm}(t)g_{lm}(r)$. Then Eq.~\eqref{S constraint} may be written as  (we drop the labels $l$ and $m$ for ease of notation)
\begin{equation}
f(t)\left[ r^2 \partial_r^2 g(r) + 2 r \partial_r g(r) - \left(l(l+1) - \omega^2 r^2\right)g(r)\right] - r^2 g(r)\left[\partial_t^2f(t) + \omega^2 f(t)\right] =0,
\end{equation}
and it splits into two ordinary differential equations for $f(t)$ and $g(r)$:
\begin{align}
\partial_t^2f(t) + \omega^2 f(t)=0,\qquad r^2 \partial_r^2 g(r) + 2 r \partial_r g(r) - \left(l(l+1) - \omega^2 r^2\right)g(r) =0.
\end{align}
There are two classes of solutions that are qualitatively different:
\begin{enumerate}
\item\textit{Oscillatory modes} ($\omega\neq 0$). In this case the solution is a linear combination of the \emph{spherical Bessel functions of the first and second kind}
\begin{equation}
S_{lm}(t,r) = e^{\pm i\omega t}\big(c_1 J_{l}( \omega r ) + c_2  Y_{l}( \omega r )\big);
\end{equation}
\item\textit{Zero modes} ($\omega= 0$). In this case the solution is expressed as linear combination of $4$ modes:
\begin{equation}
S_{lm}(t,r) = (c_1 t+c_2)r^l +(c_3 t+c_4)r^{-l-1} .
\end{equation}
\end{enumerate}
Outer symmetries are defined from the zero mode class alone. Moreover, the solution $\sim r^{-l-1}$ is a gauge mode which is discarded. We therefore focus our attention to the class of solutions given by the linear combination of the \emph{regular solid scalar harmonics} $r^l Y^{lm}$ 
\begin{equation}
\xi^{0} = (c_1 t+c_2) r^l Y^{lm} . 
\end{equation}

The coefficients $B_{lm}$ in \eqref{vector field} obey the same constraint as in Eq.~\eqref{S constraint}, because the vector harmonic $^{B}\bold{Y}^{lm}$ is an eigenvector of the operator $\bold{L}^2$. Hence, the coefficients $B_{lm}$ are again given by the above form for $S_{lm}$ and the solutions of interest read as
\begin{equation}
^{B}\xi_{i} = (c_1 t+c_2) r^l ~^{B}Y^{lm}_{i} .
\end{equation}

For the electric and the radial components of the vector field \eqref{vector field}, the harmonic gauge conditions are not satisfied independently.
Therefore, we demand that their linear combination is a solution to the D'Alembertian equation and we find four linearly independent solutions:
\begin{subequations}\label{Y1-Y4}
\begin{align}
\bold{Y}_1^{lm} &= \frac{c_1 t + c_2}{\sqrt{2l+1}} \frac{1}{r^{l+2}} \left(\sqrt{l}~^{E}\bold{Y}^{lm} - \sqrt{l+1}~^{R}\bold{Y}^{lm}  \right)= \frac{c_1 t + c_2}{\sqrt{(l+1)(2l+1)}}\nabla\left(\dfrac{Y^{lm}}{r^{l+1}}\right),\\
\bold{Y}_2^{lm} &= \frac{c_1 t + c_2}{\sqrt{2l+1}}\frac{1}{r^{l}} \left(\sqrt{l+1}~^{E}\bold{Y}^{lm} + \sqrt{l}~^{R}\bold{Y}^{lm}  \right)= \frac{c_1 t + c_2}{\sqrt{l(2l+1)}} \frac{1}{r^{2l-1}}\nabla(r^lY^{lm}),\\
\bold{Y}_3^{lm} &= \frac{c_1 t + c_2}{\sqrt{2l+1}}r^{l-1} \left(\sqrt{l+1}~^{E}\bold{Y}^{lm} + \sqrt{l}~^{R}\bold{Y}^{lm}  \right)= \frac{c_1 t + c_2}{\sqrt{l(2l+1)}} \nabla(r^lY^{lm}),\\
\bold{Y}_4^{lm} &= \frac{c_1 t + c_2}{\sqrt{2l+1}}r^{l+1} \left(\sqrt{l}~^{E}\bold{Y}^{lm} - \sqrt{l+1}~^{R}\bold{Y}^{lm}  \right)= \frac{c_1 t + c_2}{\sqrt{(l+1)(2l+1)}} r^{2l+3}\nabla\left(\dfrac{Y^{lm}}{r^{l+1}}\right) \label{Y4}.
\end{align}
\end{subequations}
Again the first two solutions are pure gauge and are discarded. The vector $\bold{Y}^{lm}_{1}$ is parallel to $\bold{Y}^{lm}_{4}$ and the vector $\bold{Y}^{lm}_{2}$ is parallel to $\bold{Y}^{lm}_{3}$. The vectors $\bold{Y}_1^{lm}$ and $\bold{Y}_3^{lm}$ are manifestly solenoidal and irrotational, because they are proportional to the gradients of solid harmonic functions. Instead, the divergence and the curl of $\bold{Y}_2^{lm}$ and $\bold{Y}_4^{lm}$ read as
\begin{subequations}
\begin{align}
\nabla \cdot \bold{Y}_2^{lm}&=-(c_1 t + c_2)(2l-1)\sqrt{\frac{l}{2l+1}}\frac{Y^{lm}}{r^{l+1}},  &\nabla \times \bold{Y}_2^{lm} &= -(c_1 t + c_2)(2l-1)\sqrt{\frac{l+1}{2l+1}}\frac{^{B}\bold{Y}^{lm}}{r^{l+1}},\\
\nabla \cdot \bold{Y}_4^{lm}&=-(c_1 t + c_2)(2l+3)\sqrt{\frac{l+1}{2l+1}}r^l Y^{lm}, &\nabla \times \bold{Y}_4^{lm} &= (c_1 t + c_2)(2l+3)\sqrt{\frac{l}{2l+1}}r^l~^{B}\bold{Y}^{lm}.
\end{align}
\end{subequations}
The vector field $\boldsymbol V$ in the main text is indeed the time independent part of $\bold{Y}_4$:
\begin{align}\label{V-vector}
\bold V^{lm}=\frac{1}{\sqrt{(l+1)(2l+1)}} r^{2l+3}\nabla\left(\dfrac{Y^{lm}}{r^{l+1}}\right)=\frac{1}{\sqrt{2l+1}}r^{l+1} \left(\sqrt{l}~^{E}\bold{Y}^{lm} - \sqrt{l+1}~^{R}\bold{Y}^{lm}  \right)
\end{align}

\section{Canonical surface charges}
\label{app:ch}

In four-dimensional General Relativity, the infinitesimal canonical \emph{surface charge} associated to a vector field $\xi$ and a linearised metric $h_{\mu\nu}$ around the background $\bar g_{\mu\nu}$ is given by
\begin{equation}
\bold{k}_{\xi}[h ; \bar g] = \sqrt{-\bar{g}} k_{\xi}^{\mu\nu}[h ; \bar g] \left(d^2x\right)_{\mu\nu},
\end{equation}
where $\left(d^2x\right)_{\mu\nu} = \frac{1}{4} \epsilon_{\mu\nu\alpha\beta}dx^{\alpha}\wedge dx^{\beta}$ and 
\begin{equation} \label{surface charge}
k_{\xi}^{\mu\nu} = \xi^{\nu} \left(\bar{\nabla}^{\mu} h - \bar{\nabla}_{\sigma} h^{\mu\sigma}\right) + \xi_{\sigma}\bar{\nabla}^{\nu}h^{\mu\sigma} + \frac{1}{2} h \bar{\nabla}^{\nu} \xi^{\mu} - h^{\rho \nu}\bar{\nabla}_{\rho}\xi^{\mu} + \frac{\alpha}{2}h^{\sigma\nu}\left( \bar{\nabla}^{\mu}\xi_{\sigma} + \bar{\nabla}_{\sigma}\xi^{\mu}\right).
\end{equation}
Here, the background metric is used to lower and to raise indices, \emph{e.g.}, $h^{\mu\nu} = \bar g^{\mu\alpha} \bar g^{\nu \beta} h_{\alpha \beta}$ and $h= \bar g^{\mu\nu} h_{\mu\nu}$. The symbol $\bar{\nabla}_{\mu}$ stands for the background covariant derivative. 

The corresponding linearised charges $Q_{\xi}[h; \bar g]$ associated to the vector field $\xi$ are obtained by integrating the spacetime 2-form $\bold{k}_{\xi}$ over a two-dimensional sphere $S$
\begin{equation} \label{charge}
Q_{\xi}[h; \bar g] = \frac{1}{8\pi G}\int_{S}\bold{k}_{\xi}[h ; \bar g]. 
\end{equation}

The parameter $\alpha$ reflects the ambiguity at adding a boundary form $\bold{E} \sim \left(d^2x\right)_{\mu\nu} (\delta g)^{\mu\alpha} \wedge (\delta g)_{\alpha}^{\; \nu}$ to the Lee-Wald symplectic form of Einstein gravity \cite{Lee:1990nz}. For $\alpha=0$, one has the Iyer-Wald charge \cite{Iyer:1994ys} and the Lee-Wald symplectic structure \cite{Lee:1990nz,Wald:1999wa}. For $\alpha=1$, one has the Abbott-Deser \cite{Abbott:1981ff} or, equivalently, the Barnich-Brandt charge \cite{Barnich:2001jy} and the invariant symplectic structure \cite{Barnich:2007bf,Compere:2007az}.

We take the background metric to be the Minkowski metric $\bar{g}_{\mu\nu} = \eta_{\mu\nu}$ and we adopt Cartesian coordinates $(t, x, y, z)=(t,x^i)$. Background covariant derivatives reduce to partial derivatives $\bar{\nabla}_{\mu} = \partial_{\mu}$. Simple algebra reduces Eq.~\eqref{surface charge} to the expression
\begin{align} \label{0i component} 
2 k_{\xi}^{[0i]} &= \xi^{i} \left( \partial_{j} h_{0j} -\dot{h} -\dot{h}_{00}   \right) + \xi^{j} \left( \dot{h}_{ij} - \partial_{i} h_{0j}\right) - \xi^{0} \left( \partial_{i} h + \partial_{i} h_{00} - \partial_{j} h_{ij}\right) + \nonumber\\
&\quad + \frac{1}{2} h \left( \partial_{i}\xi^{0} + \dot{\xi}^{i} \right) + h_{00} \dot{\xi}^{i} + h_{0i} \dot{\xi}^{0} - h_{0j} \partial_{j}\xi^{i} -h_{ij}\partial_{j}\xi^{0} +\\
&\quad + \frac{\alpha}{2} \left[h_{00} \left( \partial_{i}\xi^{0} - \dot{\xi}^{i}\right) + h_{0j} \left( \partial_{i} \xi_{j} +\partial_{j} \xi_{i}  \right) -2h_{0i} \dot{\xi}^{0} + h_{ij}\left(\partial_{j}\xi^{0} - \dot{\xi}_{j}\right)\right], \nonumber
\end{align}
where $\xi_{i} = \xi^{i}$, but $\xi_{0} = -\xi^{0}$. The dot stands for time derivative, \emph{i.e.}, $\dot{h}_{\mu\nu} = \partial_{0}h_{\mu\nu}$. For higher curvature theories or for a large class of matter theories, see e.g. \cite{Iyer:1994ys,Compere:2009dp,Azeyanagi:2009wf} for the explicit expression of $\bold{k}_{\xi}$.

\subsection{Coefficients of the surface charges}
\label{app:charges}

The coefficients appearing in \eqref{mainch1}-\eqref{mainch2}-\eqref{mainch3} are given by 
\begin{subequations}
\begin{align}
C_{L_{}}(p, l) &= -\frac{2}{(l-1)!}\sqrt{\frac{l}{l+1}}\frac{2^{p-l}}{p!} \frac{(2l-p)!}{(l-p+1)!} \times \\
& \quad\times \left[ \frac{\alpha}{2}(l-1)(l-p+1) + l(l-p+3) -2(p-1) + \frac{(p-1)p}{2}\left(1- \sqrt{\frac{2(l-1)}{l+2}}\right) \right] \nonumber,\\
C_{K_{}}(p, l) &= \frac{2}{l!}\frac{2^{p-l}}{p!} \frac{(2l-p-2)!}{(l-p+1)!} \times \\
&\times \left[\frac{1}{2}\frac{(2l-p+2)!}{(2l-p-2)!} - \sqrt{\frac{l(l-1)}{2(l+1)(l+2)}}p(p-1)\left[2+4l(1+2(l-p))+p(p-1)\right]\right] \nonumber,\\
C_{P_{}}(p, l) &= \frac{2}{l!}\frac{2^{p-l}}{p!} \frac{(2l-p-2)!}{(l-p+1)!} \Bigg\{-\frac{l (2l-p)!}{(2l-p-2)!} + (1-\delta_{1 l})\sqrt{l(l+1)}\times \\
&\hspace{-12mm} \times \left[2l-p(p+1) + \frac{2(l-p)}{l+1} \left[2(l^2-2l-1) -\alpha(2l+1)(l-1)\right] +p(p-1)\frac{2l+1}{l+1}\sqrt{\frac{2(l-1)}{l+2}}\right]\Bigg\} \nonumber.
\end{align}
\end{subequations}
For the sake of completeness, we write the charge associated to $\boldsymbol V =r^{l+1}\left(\sqrt{l} ~^{E}\bold{Y}^{lm} - \sqrt{l+1}~^{R}\bold{Y}^{lm}\right)$. It reads as
\begin{equation} \label{chargeV}
8\pi G~Q^{lm}_{\boldsymbol V} = \sum_{p=1}^{l+1}C_{\boldsymbol V}(p, l) r^{p+1}~^{(p)}I^{lm}(u),
\end{equation}
with the coefficient $C_{\boldsymbol V}(p, l)$ being
\begin{align}
&C_{\boldsymbol V}(p, l) = \frac{1}{l !} \frac{2^{p-l}}{(p-1)!}\frac{(2l-p-1)!}{(l-p+1)!} \times \\
& \times \left[\sqrt{l+1}(2l-p+1)(2l-p) - (1-\delta_{1l})\sqrt{l}\left[p(p-8l-7)+4l(2l+3)+6-(2l+3)(l-p+1)\alpha\right]\right]. \nonumber
\end{align}
From \eqref{chargeV}, we clearly see that the associated charge is vanishing at spatial infinity. Moreover, the surface integral in the near zone which could in principle correspond to a multipole moment is also vanishing. Accordingly we have ignored this vector in this paper, although it might be important in more general theories of gravity, as discussed in Section \ref{sec:discussion}. 

\subsection{Multipole charges of a harmonic gauge perturbation}
\label{chg}
The multipole charges discussed in this paper are computed in canonical harmonic gauge. As explained in section \ref{sec:gf}, the term ``canonical'' refers to the extra conditions that are met by the metric configurations in addition to respecting the harmonic or de Donder gauge. Assuming asymptotically flat boundary conditions, the canonical conditions amount to reaching mass-centered frame and reaching \eqref{canonical gauge}. The purpose of this section is to investigate whether one can relax the last condition \eqref{canonical gauge}.  

To this end, we compute  the variation of multipole charges $Q_\xi$ under a variation in the metric of the form $\de_\zeta g_\mn=\mathcal L_\zeta \eta_{\mu\nu}$\footnote{Note that $\de_\zeta  g_\mn=\mathcal L_\zeta \eta_{\mu\nu}+\mathcal L_\zeta h_{\mu\nu}$ but the latter term is vanishing in linearized theory as both $\zeta$ and $h_\mn$ are infinitesimal.} that preserves de Donder gauge, asymptotically flat boundary conditions and the mass-centered frame but violates \eqref{canonical gauge}. The most general form of such vector is given in STF harmonics in Section VIII Eq.~(8.9) of \cite{Thorne:1980ru}, while its spherical harmonic expansion turns out to be
\begin{subequations} \label{gauge transfs2}
\begin{align}
\zeta_{0} &=  \sum_{l=0}^{\infty} ~ \sum_{m=-l}^{l} (-1)^l ~\sum_{k=0}^{l}~ \frac{1}{2^k k!}\frac{(l+k)!}{(l-k)!} \left(\frac{1}{2}~^{(l-k+1)}K^{lm}(u) - ~^{(l-k)}D^{lm}(u)\right) \frac{Y^{lm}}{r^{k+1}}, \\
\zeta_{i} &= -\frac{1}{2} \sum_{l=0}^{\infty} ~ \sum_{m=-l}^{l} (-1)^l ~\sum_{k=0}^{l}~ \frac{1}{2^k k!}\frac{(l+k)!}{(l-k)!} \times \\
&\qquad \qquad \times \left[\frac{~^{(l-k)}K^{lm}(u)}{r^{k+2}} \left(\sqrt{l(l+1)}~^{E}Y^{lm}_{i}-(k+1)~^{R}Y^{lm}_{i}\right) - 
\frac{~^{(l-k+1)}K^{lm}(u)~^{R}Y^{lm}_{i}}{r^{k+1}}\right] + \nonumber\\
&\quad -\frac{1}{2}\sum_{l=1}^{\infty} ~ \sum_{m=-l}^{l} (-1)^l \Bigg[\sum_{k=0}^{l}~ \frac{1}{2^k k!}\frac{(l+k)!}{(l-k)!} \frac{~^{(l-k)}N^{lm}(u)~^{B}Y^{lm}_{i}}{r^{k+1}} + \nonumber\\
& \qquad \qquad \qquad \qquad \quad- \sum_{k=0}^{l-1}~ \frac{1}{2^k k!}\frac{(l-1+k)!}{(l-1-k)!}\frac{~^{(l-1-k)}H^{lm}(u)}{r^{k+1}}\left(^{E}Y^{lm}_{i} + \sqrt{\frac{l}{l+1}}~^{R}Y^{lm}_{i}\right)\Bigg]. \nonumber
\end{align}
\end{subequations}
The four components of the vector field $\zeta^{\mu}$ are harmonic functions and depend upon four arbitrary functions of the retarded time $u$. They must satisfy the same behavior as \eqref{CK-condition}. We now compute the charge variation $\de_\zeta Q_\xi=\frac{1}{8\pi G}\int_S \bold{k}_{\xi}[\mathcal{L}_{\zeta}\eta; \eta]$ where $\xi$ is one of the multipole symmetries \eqref{residual symm}. Explicit computation shows that the charges computed at a surface $S$ of radius $r$ are given by

\begin{subequations}\label{charge variation-gauge}
\begin{align}
8\pi G~\de_\zeta Q^{lm}_{L} &= (-1)^l (l-1)\sqrt{l(l+1)}~ \frac{\alpha}{8}~\sum_{p=0}^{l} \frac{2^{p-l}}{p!} \frac{(2l-p)!}{(l-p)!} ~r^p ~^{(p+1)}N^{lm},\\
8\pi G~\de_\zeta Q^{lm}_{K} &= -(-1)^l (l-1) l ~\sum_{p=0}^{l} \frac{2^{p-l}}{p!} \frac{(2l-p-2)!}{(l-p-1)!} r^{p} \Bigg[ \frac{2l+1}{\sqrt{l(l+1)}} ~^{(p)}H^{lm} + ^{(p+2)}K^{lm} \Bigg] \nonumber \\
&\quad+ 8\pi G\, t \, \de_\zeta Q^{lm}_{P_{\epsilon}} , \\
8\pi G~\de_\zeta Q^{lm}_{P} &= (-1)^l (l-1) l ~\sum_{p=0}^{l} \frac{2^{p-l}}{p!}  \frac{(2l-p-2)!}{(l-p-1)!}\times  \\ 
&\qquad \times r^{p} \left( (\alpha-2) ^{(p+2)}D^{lm} + ^{(p+3)}K^{lm} + \frac{\alpha}{2} \frac{2l+1}{\sqrt{l(l+1)}} ~^{(p+1)}H^{lm}\right). \nonumber
\end{align}
\end{subequations}
Now we send $S$ to spatial infinity. As a result, all the charge variations vanish, except the mass multipole charge variation $\de_\zeta Q^{lm}_{K}-t \de_\zeta Q^{lm}_{P}$, due to the function $H^{lm}$. These gauge transformations correspond to functions $\zeta^i = \p_i \eps$ with $\eps$ harmonic. 

For the sake of completness, the charge associated to $\bold{V}^{lm} = r^{l+1}\left(\sqrt{l} ~^{E}\bold{Y}^{lm} - \sqrt{l+1}~^{R}\bold{Y}^{lm}\right)$ is given by
\begin{align}
8\pi G~\de_\zeta Q^{lm}_{\bold{V}} &= \frac{1}{2}(-1)^l\sqrt{l+1}\Biggl\{\sum_{p=0}^{l+1}\frac{2^{p-l}}{p!}\frac{(2l-p+1)!}{(l-p+1)!} r^p \Biggl[d(p,l)~^{(p)}D^{lm}(u) + k(p,l)~^{(p+1)}K^{lm}(u)\biggr]+\nonumber \\
\qquad & +\sum_{p=1}^{l+1}\frac{2^{p-l}}{p!}\frac{(2l-p+1)!}{(l-p+1)!} h(p,l)r^p ~^{(p-1)}H^{lm}(u)\Biggr\},
\end{align}
where
\begin{subequations}
\begin{align}
d(p,l) &= -\frac{1}{2(2l-p+1)} \Big\{ 2+4l^3+2l(6-5p)+2l^2(7-2p)+p(p-3)+\\
\quad &\hspace{2cm}+\alpha[2+2l^3+l^2(5-2p)+l(p^2-4p+5)+p(p-3)]\Big\} \nonumber,\\
k(p,l)&=\frac{1}{2}(l+2)[2(l+1)-p],\\
h(p,l)&=\frac{1}{4}\sqrt{\frac{l}{l+1}}(2l+3)\left(1-\frac{\alpha}{2}\right)\frac{p(p-1)}{2l-p+1}.
\end{align}
\end{subequations}

\newpage
\bibliography{ref_multipoles}

\end{document}